\documentclass[11pt,letterpaper]{article}
\usepackage[top=1in,left=1in,right=1in,bottom=1in]{geometry}

\usepackage{amsmath,amssymb}

\usepackage{changepage}

\usepackage[utf8x]{inputenc}

\usepackage{textcomp,marvosym}

\usepackage{cite}

\usepackage{nameref,hyperref}

\usepackage{microtype}
\DisableLigatures[f]{encoding = *, family = * }

\usepackage[table]{xcolor}

\usepackage{array}

\RequirePackage{amsthm,amsmath,amsfonts,amssymb}

\usepackage{amsmath}
\usepackage{graphicx}
\usepackage{etoolbox}

\usepackage{amsfonts}            
\usepackage{amssymb}             

\usepackage{amsthm}              
\usepackage{amsmath}            
\usepackage{algorithm}           
\usepackage{algorithmic}
\usepackage{stackrel}
\usepackage{float}
\usepackage{placeins}
\usepackage{dblfloatfix}
\usepackage{threeparttable,array,booktabs}
\usepackage{changepage}
\usepackage{url}
\usepackage{bm,bbm}

\newcolumntype{+}{!{\vrule width 2pt}}

\newlength\savedwidth

\usepackage[aboveskip=1pt,labelfont=bf,labelsep=period,justification=raggedright,singlelinecheck=off]{caption}

\makeatletter
\renewcommand{\@biblabel}[1]{\quad#1.}
\makeatother

\usepackage{lastpage,fancyhdr,graphicx}
\usepackage{epstopdf}
\fancyheadoffset[L]{2.25in}
\fancyfootoffset[L]{2.25in}

\newcommand{\citep}[1]{{\cite{#1}}}
\newcommand{\citet}[1]{{\cite{#1}}}

\DeclareMathOperator*{\argmax}{arg\,max}

\begin{document}
\vspace*{0.2in}

{\Large
\textbf\newline{Highly Scalable Maximum Likelihood and Conjugate Bayesian Inference for ERGMs on Graph Sets with Equivalent Vertices} 
}
\newline
\\
Fan Yin\textsuperscript{1},
Carter T. Butts\textsuperscript{2\ddag}
\\
\bigskip
\textbf{1} Department of Statistics, University of California at Irvine, Irvine CA, USA
\\
\textbf{2} Departments of Sociology, Statistics, Computer Science, and EECS
and Institute for Mathematical Behavioral Sciences, University of California at Irvine, Irvine, CA, USA
\\
\bigskip

%



\ddag Corresponding author. Email: \texttt{buttsc@uci.edu}



\section*{Abstract}
The exponential family random graph modeling (ERGM) framework provides a highly flexible approach for the statistical analysis of networks (i.e. graphs). As ERGMs with dyadic dependence involve normalizing factors that are extremely costly to compute, practical strategies for ERGMs inference generally employ a variety of approximations or other workarounds. Markov Chain Monte Carlo maximum likelihood (MCMC MLE) provides a powerful tool to approximate the maximum likelihood estimator (MLE) of ERGM parameters, and is generally feasible for typical models on single networks with as many as a few thousand nodes. MCMC-based algorithms for Bayesian analysis are more expensive, and high-quality answers are challenging to obtain on large graphs. For both strategies, extension to the pooled case - in which we observe multiple networks from a common generative process - adds further computational cost, with both time and memory scaling linearly in the number of graphs. This becomes prohibitive for large networks, or cases in which large numbers of graph observations are available. Here, we exploit some basic properties of the discrete exponential families to develop an approach for ERGM inference in the pooled case that (where applicable) allows an arbitrarily large number of graph observations to be fit at no additional computational cost beyond preprocessing the data itself. Moreover, a variant of our approach can also be used to perform Bayesian inference under conjugate priors, again with no additional computational cost in the estimation phase. The latter can be employed either for single graph observations, or for observations from graph sets. As we show, the conjugate prior is easily specified, and is well-suited to applications such as regularization. Simulation studies show that the pooled method leads to estimates with good frequentist properties, and posterior estimates under the conjugate prior are well-behaved.  We demonstrate the usefulness of our approach with applications to pooled analysis of brain functional connectivity networks and to replicated x-ray crystal structures of hen egg-white lysozyme.





\section{Introduction}

Networks are relational structures composed of individual entities (\emph{vertices} or \emph{nodes}) together with a set of pairs or ordered pairs of entities (\emph{ties} or \emph{edges}) that share a specific relationship. Networks arise in many scientific fields, ranging from biology and epidemiology to social science and engineering. For example, social science researchers are frequently interested in interpersonal networks, in which nodes correspond to individuals and edges represent personal relationships (e.g. friendship \citep{smith2016ethnic}, advice-seeking \citep{cross.et.al:sn:2001}, etc.); in biology, there has been research interest in using networks to represent complex phenomena such as transcriptional regulation \citep{saul2007exploring}, trophic systems \citep{SAINTBEAT2015458}, interspecific competition \citep{delmas.et.al:br:2019}, animal social interaction \citep{krause.et.al:bk:2015},  protein structure \citep{cross.et.al:bio:2020} and aggregation \citep{grazioli.et.al:jpcB:2019}, and the structure and function of neural systems \citep{cook.et.al:n:2019}. As these disparate examples illustrate, networks have proven to be a fruitful framework for treating a wide range of phenomena, and research on network structure has grown apace. 

Accompanying this growth has been a corresponding literature on techniques for network measurement, modeling, and analysis.  Recent advances in inferential methods for models with complex dependencies have enabled the statistical modeling and analysis of network data to become a practical tool for a growing range of research applications \citep{kolaczyk:bk:2009,snijders2011statistical,salter2012review}. A variety of modeling approaches have benefited from these advances, particularly exponential family random graph models (ERGMs) \citep{lusher.et.al:bk:2012}, parametric model families that are capable of capturing the complex dependence structure that is typical of network data. ERGMs (known in older work as $p*$ models, e.g. \citep{wasserman1996logit}) gradually developed from early tools for testing dependence hypotheses to a general framework for modeling networks with heterogeneity and complex dependence \citep{holland1981exponential, frank1986markov, snijders2006new, pattison.robins:sm:2002,hunter2006inference}, and have spawned a growing body of theoretical \citep{strauss1986general,haggstrom.jonasson:jap:1999,handcock:ch:2003,rinaldo2009geometry,schweinberger2011instability,chatterjee.diaconis:aos:2013,butts:jms:2019,butts:jms:2020b} and methodological work \citep{koskinen2004bayesian,caimo2011bayesian,hunter2012computational,hummel2012improving,krivitsky2012exponential,koskinen.et.al:sn:2013,kolaczyk2015question}. (See \citep{schweinberger.et.al:ss:2020} for a recent review.)  ERGMs have been widely applied in many scientific fields, for example, including (but not limited to) sociology \citep{goodreau2009birds,smith2016ethnic, srivastava2011culture}, political science \citep{cranmer2011inferential}, bioinformatics \citep{saul2007exploring}, public health \citep{welch2011statistical}, biophysics \citep{grazioli.et.al:fmb:2019,grazioli.et.al:jpcB:2019}, and neuroscience \citep{simpson2011exponential, simpson2012exponential}.

Research on statistical network models has been particularly concerned with the case of inference from a single network observation, but multiple random network realizations are also increasingly common in practice, and can be divided into major two categories depending upon the nature of the underlying vertex set. The first category involves cases such as networks from independently solved protein structures or molecular dynamics (MD) simulations, brain functional connectivity networks \citep{simpson2011exponential,simpson2012exponential,sinke2016bayesian}, or dynamic friendship networks within fixed groups \citep{knecht2008friendship}, where the node sets on which the relationship is defined are constant across observations. There are also cases where node sets across different observations are potentially non-equivalent, including comparative studies of networks across groups \citep{zijlstra2006multilevel} or species \citep{faust.skvoretz:sm:2002}, intervention studies for education research \citep{sweet2013hierarchical, sweet2014hierarchical} and friendship network studies across different schools \citep{goodreau2009birds}, or studies of emergent multi-organizational networks with changing composition \citep{butts.et.al:joss:2012}.  A second useful distinction involves cases in which multiple independent (or approximately independent) networks are observed (e.g., from populations of subjects or organizations, or from well-spaced snapshots of equilibrium dynamics), versus time series data in which draws are strongly autocorrelated.  The latter is heavily studied as a case in its own right (e.g., \citep{snijders:sm:2001,koskinen.snijders:jspi:2007,hanneke2010discrete, desmarais2012statistical,almquist2014bayesian,krivitsky2014separable}), with a primary focus on uncovering the mechanisms governing network dynamics.  Models for the former case generally seek to either strictly summarize structural variation in a population of graphs (e.g., \citep{faust.skvoretz:sm:2002,butts:sm:2011a}) or pool information in a common population model (e.g., \citep{koehly.pattison:ch:2005,zijlstra2006multilevel,sweet2013hierarchical, sweet2014hierarchical}) that infers common structural tendencies in a more generative fashion.  As we detail below, our focus in this paper is on models for the equivalent vertex set/non-autocorrelated pooled case.

Despite considerable progress in this area, there remains room for improvement in approaches for inference based on multiple network observations with ERGMs. Broadly speaking, existing approaches based on full likelihood calculations (as opposed to composite/pseudo-likelihood methods, e.g. \citep{faust.skvoretz:sm:2002,cranmer2011inferential,grazioli.et.al:fmb:2019}) can be divided into two categories, unpooled and pooled estimates.  Unpooled estimation employs an essentially meta-analytic approach, in which a proposed model is fit to each network observation separately, with the resulting parameter estimates being jointly analyzed in a second stage.  For example, motivated by the goal of finding a model for brain-connectivity networks at the group level, \citet{simpson2012exponential} specified an ERGM model for brain-connectivity networks and then fit the model to each individual separately, subsequently combining the resulting estimates by taking their respective means and medians as point estimates of the parameters of a group-level representative model. A similar approach was used by \citet{goodreau2009birds} in studying friendship across schools, with separate models fit to each network and the resulting statistics then summarized to infer general patterns. Under such a framework, the time complexity of inference scales linearly in the number of graphs, making this generalization expensive as the number of networks grows (particularly if the individual networks are themselves large). A second problem with the unpooled approach is that it may be difficult or impossible to find a model that is both estimable on each individual network and that includes all effects of substantive interest.  For instance, where the model sufficient statistics for a particular network are sufficiently extreme (in a sense to be clarified below), the MLE may not exist; this condition is common for effects involving subgroup interactions when said subgroups are small.  Importantly, this failure need not mean that the model family is generally inappropriate or ill-behaved, instead stemming from limitations in the ability to fit some models to a single graph realization.  A natural way to avoid such problems is by pooled estimation.  In the ERGM context, pooled estimation has been studied e.g. in the case of independently observed intraschool friendship networks \citep{STEWART201998}, where the adjacency matrices representing the observed networks of each distinct school are aggregated to a block-diagonal matrix with structural zeros assumed for all off-diagonal blocks. However, the high cost of performing MCMC MLE on such a pooled network greatly limits the scale of cases that can be considered in this way. Similar schemes using hierarchical Bayesian models have also been proposed \citep{slaughter2016multilevel}, but require high-quality MCMC simulations that can be computationally demanding when the number of network observations is large.  Importantly, all of these methods share the property that likelihood calculations must \emph{de facto} be performed for each network separately (whether those networks are notionally joined together in one large synthetic network as part of a pooling scheme or treated separately), which greatly increases both storage and time complexity (especially for large graphs).  

This cost poses a substantial barrier in applications such as neuroscience or biophysics, where pooled inference on large collections of networks is of potential interest.  Importantly, however, some of these cases have the special property that the collections of networks to be analyzed (or large subsets thereof) involve equivalent vertex sets (with either absent or equivalent covariates).  For instance, \citet{grazioli.et.al:fmb:2019} study two collections of approximately 1,000 networks representing independently drawn local energy minima for structures of wild type and E22G variants of A$\beta_{1-40}$, a protein that plays a key role in the etiology of Alzheimer's disease.  Each collection represents a series of independent draws from a respective common graph distribution, with identical vertex properties for graphs in each set.  Likewise, in neuroscience settings one may (as in e.g. \citep{simpson2011exponential}) observe anatomically defined networks on collections of subjects that are (at least provisionally) considered exchangeable within groups, and that can be usefully modeled as draws from a common graph generating process.  Although less common to date in social science settings, collections of exchangeable networks with equivalent vertex sets can arise e.g. from behavior in human subject experiments \citep{VEGAYON2021225}, semantic networks extracted experimentally or from texts \citep{zemla.austerweil:cbb:2018}, or as the result of replicated outcomes of agent-based simulations \citep{beskow.carley:pr:2019}.  In such cases, it is possible to perform pooled ERGM inference at vastly lower cost than is possible with conventional techniques - indeed, obtaining computational costs for estimation that are identical to the single-graph case.

In this paper, we propose a novel method for fitting multiple graph observations embedded in equivalent node sets using a pooled ERGM approach.  By exploiting a simple property of statistical exponential families, we are able to convert the problem of pooled inference for a (possibly large) graph set to the problem of inference for a single pseudo-graph of size equal to an individual input graph, subsequently correcting the single graph information matrix for the sample size of the pooled data set.  We also show that a minor adjustment to this technique can be used to perform maximum \emph{a posteriori} (MAP) estimation under conjugate priors at no additional cost, either for a single graph or for multiple graph observations.  Because our technique works entirely within the \emph{mean value space} of the chosen ERGM family, it can be performed via data adjustments with existing software intended for single-graph estimation, and is compatible with any estimation method that works via sufficient statistics (including the widely used Geyer-Thompson \citep{geyer1992constrained,hunter2008ergm} and stochastic approximation \citep{snijders2002markov} methods).  In addition to pooling information and providing inexpensive Bayesian answers (using the Laplace approximation to the posterior distribution), our approach provides a simple and effective mechanism for regularization, with particular virtue in resolving the ``convex hull problem'' that frequently arises in discrete exponential families; we describe a simple approach to prior specification that is well-suited to this purpose, and that can be easily extended to provide more informative priors when appropriate background information is available. 

The remainder of this paper is organized as follows. We begin with general concepts and notation for ERGMs in Section~\ref{sec_ergm}, and present our framework for scalable inference under both frequentist and Bayesian settings (including issues of prior specification) in Section~\ref{sec_pooling}. In Section~\ref{sec_sim}, we employ simulation studies to examine the performance of our approach, and the behavior of posterior inference as a function of prior weight. In Section~\ref{sec_apps}, our proposed methods are demonstrated on two different multiple network applications: brain functional connectivity networks from multiple subjects (Section~\ref{sec_brain}); and protein structure networks from replicated x-ray crystal structures of hen egg-white lysozyme (Section~\ref{sec_protein}). Finally, we close with a brief discussion and comment on potential future work.

\subsection{Exponential Family Random Graph Models} \label{sec_ergm}

Consider an order-$n$ graph, $G$, represented via adjacency matrix $Y$ on support $\mathcal{Y}_n$, such that $Y_{ij}$ corresponds to the state of the edge between vertices $i$ and $j$; we make no particular assumptions about $\mathcal{Y}_n$ (e.g., it may consist of directed or undirected graphs, with or without loops, and may be valued), save that all elements of $y \in \mathcal{Y}_n$ are real and finite.  An \emph{exponential family random graph model} (ERGM) for $Y$ is then given by
\begin{equation} 
\label{eq:ERGM}
P_{\eta}(Y=y) = h(y) \exp\left\{ \eta(\theta)^\intercal g(y) - \psi( \eta(\theta) ) \right\} 
\end{equation}
where $g : \mathcal{Y}_{n} \rightarrow \mathbb{R}^{p}$ is a vector of real-valued sufficient statistics capturing network features of interest  (which may implicitly incorporate e.g., nodal or dyadic covariates) and $\theta \in \Theta \subseteq \mathbb{R}^{q}$ is vector of (curved) model parameters mapped to canonical parameters $ \eta : \theta \rightarrow \mathbb{R}^{p} $ \citep{hunter2006inference}. The reference measure $h$ determines the baseline behavior of the ERGM distribution when $\eta(\theta)=0$, and plays an important role in fixing the shape of the distribution when edges are valued \cite{krivitsky2012exponential}. In general, computation involving \eqref{eq:ERGM} is challenging due to the intractable nature of the log-partition function (i.e. normalizing factor),
$\psi(\eta(\theta)) = \log \sum_{y' \in \mathcal{Y}_{n} } h(y') \exp\left\{ \eta(\theta)^\intercal g(y') \right\}$, as $|\mathcal{Y}_n|$ is extremely large ($\mathcal{O}(2^{n^2})$ in the binary case), the summand is generally too rough for naive Monte Carlo strategies to converge well, and $\psi$ rarely has a closed-form solution.  In the context of iid draws from the same ERGM pmf, we obtain the (homogeneous) \emph{pooled ERGM}, 
\begin{equation} 
\label{eq:iidERGM}
P_{\eta}(\mathbf{Y}=\mathbf{y}) = \exp\left\{ \eta(\theta)^\intercal \sum_{i=1}^m g(y^i) +\sum_{i=1}^m \log h(y^i)- m \psi( \eta(\theta) ) \right\}, 
\end{equation}
where $\mathbf{Y}=(Y^1,\ldots,Y^m)$ is a vector of random graphs with realizations $\mathbf{y}=(y^1,\ldots,y^m)$.  Although most work focuses on the single-graph case, our emphasis here is on the case where $m>1$ (either because of multiple graph observations, or - in the case of conjugate prior inference - because of ``effective'' prior observations that are equivalent to an increased $m$).

As exponential families, the ERGMs have a number of convenient properties of which we will make use \citep{schweinberger.et.al:ss:2020}.  Subject to mild regularity conditions, we may define an invertible function $\mu(\eta) = \mathbb{E}_\eta g(Y)$ that provides the \emph{mean value parameterization} of an ERGM on random graph $Y$.  From Eq.~\ref{eq:iidERGM} it is evident that the corresponding function $\mu_m(\eta) = \mathbb{E}_\eta g(\bm{Y}) = m \mu(\eta)$ is simply a constant multiple of the base mean value function for a single graph (foreshadowing a property that we employ below).  Likewise, the Fisher information matrix of $Y$ is given by $\bm{I}(\eta)=\mathbb{E}_\eta \left[\nabla \ln P_{\eta}(\mathbf{Y}) (\nabla \ln P_{\eta}(\mathbf{Y}))^{\intercal}\right] = \mathrm{Var}_\eta g(Y)$, with the pooled equivalent being $\bm{I}_m(\eta)=m \bm{I}(\eta)$.

\section{Mean Value Inference for Pooled ERGMs} \label{sec_pooling}

Although a number of variants exist, standard approaches to inference for pooled ERGMs share the basic approach of computing likelihoods (or in some cases pseudo-likelihoods) for all observed graphs, and using the resulting joint likelihood for inference.  Computationally, this may involve (as e.g. in \citep{STEWART201998}) combining the observed graphs into a single large synthetic network of order $|V|=mn$ (with support constraints prohibiting cross-graph ties), and then performing MCMC MLE or comparable Bayesian analyses on the synthetic graph; for pseudo-likelihood methods (e.g., \citep{strauss1990pseudolikelihood,hunter2012computational}), edge variables may simply be combined across networks, possibly with resampling over networks (as with bootstrap \citep{desmarais2012statistical,schmid2017exponential} or Bayesian bootstrap \citep{grazioli.et.al:fmb:2019} strategies).  These strategies lead to computational and storage costs that increase at least linearly in the number of graphs, which can become prohibitive for large systems or when the number of graphs is substantial.  Here, we observe that a much faster strategy based on the mean values of the sufficient statistics becomes available in the IID case, and that this same strategy can also be leveraged for conjugate Bayesian inference.  To our knowledge, this very simple but powerful trick has not previously been exploited in the ERGM context.

\subsection{Maximum Likelihood Inference for Pooled ERGMs} \label{sec_mle}

We begin with the simple case of maximum likelihood inference.  Given IID ERGM observations $\bm{y}^{obs} = ( y^1, y^2, \cdots, y^m$), the joint log-likelihood follows immediately from Eq.~\ref{eq:iidERGM},
\begin{equation}\label{eq:jntlik}
\ell(\theta;\bm{y}^{obs}) =  \eta(\theta)^\intercal \sum_{i=1}^m g(y^i) + \sum_{i=1}^m \log h(y^i) - m \psi( \eta(\theta) ),
\end{equation}
the maximizer of which ($\hat{\theta}=\arg\max_\theta \ell(\theta;\bm{y}^{obs})$) is the maximum likelihood estimator (MLE).  As observed, the primary challenge in finding the MLE is in dealing with the log normalizing factor, $\psi$.  Running a Markov chain over the states of each of the $m$ graphs in the set can be used to accomplish this, or equivalently (as is done in e.g. \citet{STEWART201998}) running a single Markov chain on a combined graph of order $nm$ containing the union of all individual graphs, but the form of Eq.\ref{eq:jntlik} shows that this is superfluous in the IID case.  Specifically, observe that any maximizer of $\ell$ is also a maximizer of any positive constant multiple of $\ell$, and thus
\begin{align*}
\hat{\theta} &= \arg\max_\theta \ell(\theta;\bm{y}^{obs})\\
&= \arg\max_\theta \ell(\theta;\bm{y}^{obs})/m\\
&= \arg\max_\theta \eta(\theta)^\intercal \bar{g}(\bm{y}^{obs}) + \log \widetilde{h}(\bm{y}^{obs}) -  \psi( \eta(\theta) ),
\end{align*}
where $\bar{g}(\bm{y}^{obs})=\tfrac{1}{m}\sum_{i=1}^m g(y^i)$ is the arithmatic mean of the observed statistics, and $\widetilde{h}(\bm{y}^{obs})=\exp\left[\tfrac{1}{m} \sum_{i=1}^m \log h(y^i)\right]$ is the geometric mean of the reference measure over the observed graphs.  Since the latter does not depend on $\theta$, we may further simplify the above to
\begin{equation}
\hat{\theta} = \arg\max_\theta \eta(\theta)^\intercal \bar{g}(\bm{y}^{obs}) -  \psi( \eta(\theta) ), \label{eq:mleopt}
\end{equation}
which is immediately recognizable as the MLE for a hypothetical \emph{single} ``pseudo-graph'' of order $n$ with whose statistics are the means of the observed statistics.  It is thus possible to find the MLE for a pooled model on $m$ graphs by fitting a single-graph model (a considerable simplification).

To see the corresponding implications for the sampling distribution of the MLE, we note that inference for $\theta$ benefits from standard asymptotics in $m$ (see e.g., \citep{schweinberger.et.al:ss:2020}), including the consistency and asymptotic normality of the MLE under suitable regularity conditions.  In particular, if $\hat{\theta}_m$ is the MLE for $\mathbf{Y}$ with $m$ observations drawn from a pooled ERGM with parameter $\theta_0$, then it follows from standard exponential family theory \citep{efron:as:1975} that
\begin{equation} \label{eq:ERGMasymp}
\sqrt{m}(\hat{\theta}_m-\theta_0) \xrightarrow{\mathcal{D}} \mathcal{N}(0,\bm{I}^{-1}(\theta))
\end{equation}
where $\bm{I}(\theta)$ can be obtained from $I(\eta)$ via the chain rule, i.e. $\nabla \eta(\theta)^{\intercal} I(\eta) \nabla \eta(\theta)$.   It thus follows that the asymptotic variance-covariance matrix of the MLE in the $m$-graph case is equal to that of the single-graph case, divided by $m$; i.e.
\begin{equation}
\mathrm{Var} \hat{\theta}_m \to \tfrac{1}{m} \nabla \eta(\theta)^{\intercal} \left[\mathrm{Var} g(Y) \right]\nabla \eta(\theta).
\end{equation}

It follows, then, that we may perform maximum-likelihood inference for $\bm{y}^{obs}$ with arbitrarily large $m$ \emph{at no greater cost than fitting to a single network} (and without the use of customized software tools): we simply find the MLE for the a single (imaginary) graph with statistics equal to the mean of the observed statistics using any standard method (e.g., MCMCMLE \citep{hunter2012computational} or stochastic approximation \citep{snijders2002markov}), and then rescale the associated variance-covariance matrix by a factor of $m$ to correct for sample size.  This procedure is summarized in Algorithm~\ref{alg:pooled_ergm}.  When $m$ is large, this can result in considerable computational savings; although the trick is quite trivial to implement, it has not to our knowledge been employed for ERGM inference in prior work.

\begin{algorithm}
\caption{Maximum Likelihood Inference for a Pooled ERGM Using Mean Values \label{alg:pooled_ergm}}
\begin{algorithmic}[1]
\REQUIRE Observed data $\bm{y}^{obs}$ 
\STATE Compute $\bar{g}(\bm{y}^{obs}) =\tfrac{1}{m} \sum_{i=1}^m g(y_i)$
\STATE Find $\hat{\theta} = \arg\max_\theta \eta(\theta)^\intercal \bar{g}(\bm{y}^{obs}) -  \psi( \eta(\theta) )$
\STATE Find $\widehat{\mathrm{Var}}  \hat{\theta_m} = \tfrac{1}{m} \left[\nabla \eta(\hat{\theta})^{\intercal} I(\hat{\theta}) \nabla \eta(\hat{\theta})\right]^{-1}$
\ENSURE $\hat{\theta_m}, \widehat{\mathrm{Var}}  \hat{\theta_m} $ 
\end{algorithmic}
\end{algorithm}

\subsection{Conjugate Maximum-a Posteriori Inference for ERGMs} \label{sec_map}

We now consider the problem of IID pooled ERGM inference in the Bayesian setting. Given prior $\pi(\theta)$, we are interested in the posterior distribution of $\theta$,  $\pi(\theta | \bm{y}^{obs})$, 
\begin{align}
\pi(\theta | \bm{y}^{obs}) & = \frac{P_{\theta}(Y^{1} = y^i, Y^{2} = y^2, \cdots, Y^{m} = y^m) \pi(\theta)}{\int P_{\theta}(Y^{1} = y^1, Y^{2} = y^2, \cdots, Y^m = y^m) \pi(\theta) d\theta} \nonumber \\
& \propto \pi(\theta) \prod_{i=1}^{m} P_{\theta}(Y^{i} = y^i). \label{eq:ergm_bayes_joint}
\end{align}
Our focus here is on conjugate priors, in the canonical exponential family context for which $\eta(\theta)=\theta$.  In addition to their mathematical convenience, conjugate priors are attractive in the context of exponential families due to their interpretability (being able to be expressed in terms of prior ``pseudo-data,'' consisting of a prior ``mean''  and effective ``sample size'' expressed in the same units as the observed data), the fact that they admit natural non-informative limits, and their status as maximum entropy distributions \citep{jaynes:bk:1983}.  To our knowledge, conjugate priors for ERGMs were first examined in the unpublished work of \citet{wang:bk:2011}, who considered them along with a number of other ERGM prior specifications, but to date they have not been extensively studied.  As we show, ERGM conjugate priors allow for extremely computationally efficient inference via their mean value representation.  Moreover, there are particularly natural choices of weakly informative conjugate priors that are well-suited to regularization; we consider these in section~\ref{sec_priors}. 

For a canonical ERGM family with $\eta(\theta)=\theta$, conjugate priors take the following form \citep{almquist2014bayesian}:
\begin{equation}
\label{eq:ergm_conjugate_prior}
\pi(\theta | \bar{\tau}, n_{0}) = H(\bar{\tau}, n_{0}) \exp\left\{ n_{0} \bar{\tau}^{\intercal} \cdot \theta - n_{0}\psi(\theta) \right\}.
\end{equation}
Here, $\bar{\tau}$ are prior expected values of the vector of sufficient statistics, and $n_{0}$ is a positive number that measures the confidence in those prior expectations, which can be viewed as the number of \emph{pseudo-observations}' worth of information (in units of observed graphs) contained in the prior; $H(\bar{\tau}, n_{0})$ denotes the normalizing factor that makes $\pi(\theta | \bar{\tau}, n_{0})$ a legitimate probability density function of $\theta$. The existence of such distribution is ensured by \citet{diaconis1979conjugate}, who showed \eqref{eq:ergm_conjugate_prior} is normalizable provided that $n_0 > 0$ and $\bar{\tau}$ lies in the interior of convex hull of the support of the measure $\theta$. Substitute \eqref{eq:ergm_conjugate_prior} for $\pi(\theta)$ in \eqref{eq:ergm_bayes_joint}, we have
\begin{align}
\pi(\theta | \bm{y}^{obs}, \bar{\tau}, n_0)  & \propto \exp\left\{ [n_{0} \bar{\tau} + \sum_{i=1}^{m}  g(y^i ) ]^{\intercal} \theta - 
 (n_{0} + m) \psi(\theta)\right\} \nonumber \\
 & = \exp\left\{ [ \frac{n_{0}\bar{\tau} + \sum_{i=1}^{m}  g(y^i) }{n_0 + m} ]^{\intercal} \theta - 
 \psi(\theta)\right\}^{n_0 + m} \nonumber \\
 & = \exp\left\{ [ \delta \bar{\tau} + (1 - \delta)\bar{g}(\bm{y}^{obs}) ]^{\intercal} \theta - 
 \psi(\theta)\right\}^{n_0 + m},  \label{eq:ergm_conjugate_posterior}
\end{align}
where $\delta \equiv \frac{n_0}{n_0 + m}$, taking values in $[0,1]$. With \eqref{eq:ergm_conjugate_posterior}, we note that an analytical form for the prior $\pi(\theta | \bar{\tau}, n_0 )$ is not necessary for Bayesian inference, because the prior can be fully characterized by $\delta$ (or $n_0$) and $\bar{\tau}$. Standard Bayesian theory tells us that the posterior expectation of $\nabla \psi(\theta)$ is the Bayes estimate of $\theta$ with respect to quadratic loss \citep{bernardo2001bayesian}, and is a weighted average of $\bar{g}(\bm{y}^{obs})$ and $\bar{\tau}$, with $\delta$ controlling the relative weight of contribution of the prior information. For any given prior hyperparameters $(\bar{\tau}, n_{0})$, as the sample size $m$ becomes large, $\delta$, the relative weight of prior, approaches zero, and hence the sample-based information dominates the posterior. 

Given a prior specified by $\delta$ and $\bar{\tau}$, the maximum \emph{a posteriori} probability (MAP) estimate $\hat{\theta}^{MAP}_{m, \delta}$, is also Bayes estimate under a different choice of loss function (the 0-1 loss; see e.g. \citep{bernardo2001bayesian}). Note that since MAP is indeed the maximizer of the kernel of posterior density \eqref{eq:ergm_conjugate_posterior}, we can employ the same arguments as in the derivation of \eqref{eq:mleopt}, to obtain
\begin{align}
\hat{\theta}^{MAP}_{m, \delta} & = \argmax\limits_{\theta}  \exp\left\{ [ \delta \bar{\tau} + (1 - \delta)\bar{g}(\bm{y}^{obs})   ]^{\intercal} \theta - 
 \psi(\theta)\right\}^{n_0 + m} \nonumber \\
 & = \argmax\limits_{\theta} \exp\left\{ [ \delta \bar{\tau} + (1 - \delta)\bar{g}(\bm{y}^{obs})  ]^{\intercal} \theta - 
 \psi(\theta)\right\} \nonumber \\
 & = \argmax\limits_{\theta} \ [ \delta \bar{\tau} + (1 - \delta)\bar{g}(\bm{y}^{obs})  ]^{\intercal} \theta - 
 \psi(\theta).  \label{eq:map}
\end{align}
It follows, then, that the pooled ERGM MAP estimator $\hat{\theta}^{MAP}_{m, \delta}$ is equal to the MLE $\hat{\theta}$ that would be obtained for a single pseudo-observation with sufficient statistics $\delta \bar{\tau} + (1 - \delta)\bar{g}(\bm{y}^{obs})$.

Under standard regularity conditions, the posterior distribution $\pi(\theta | \bm{y}^{obs}, \bar{\tau}, n_0)$ becomes asymptotically Gaussian as $m\to\infty$, according to the classical Berstein-von Mises theorem \citep{van2000asymptotic}. Following the same basic ``mean value'' procedure used in Algorithm~\ref{alg:pooled_ergm} for obtaining the pooled ERGM MLE $\tilde{\theta}_{m}$, we are able to compute the MAP estimate $\tilde{\theta}^{MAP}_{m, \delta}$ by fitting an ERGM to a single 'pseudo'-graph whose node set is the same as the observed networks but whose network statistics are taken to be equal to $\delta \bar{\tau} + (1 - \delta)\bar{g}(\bm{y}^{obs})$. In addition to the MAP estimate, we can also obtain an estimate of the observed Fisher information $\hat{\bm{I}}(\tilde{\theta}^{MAP}_{m, \delta})$, which is approximately the negative Hessian of log-posterior generated by product of the prior and the likelihood of a single 'pseudo'-graph. However, the Laplace approximation of posterior distribution requires the Hessian of true log-posterior, which should be generated as by the product of prior and the likelihood of all actual observations. Note that the negative Hessian matrix $Q_{m,\delta}(\theta)$ of true log-posterior \eqref{eq:ergm_conjugate_posterior} can be approximated by $Q_{m,\delta}(\tilde{\theta}^{MAP}_{m, \delta}) \approx (m + n_{0})\hat{\bm{I}}(\tilde{\theta}^{MAP}_{m, \delta})$ = $\frac{m}{1-\delta} \hat{\bm{I}}(\tilde{\theta}^{MAP}_{m, \delta})$. \emph{Laplace's approximation} of the posterior yields \citep{tierney1986accurate} the following result,  
\begin{equation}
\label{eq:laplace_approximation}
\theta | \bm{y}^{obs} \overset{\cdot}{\sim} \mathcal{N} \big( \ \tilde{\theta}^{MAP}_{m, \delta}, \ Q_{_{m,\delta}}^{-1}(\tilde{\theta}^{MAP}_{m, \delta}) \big).
\end{equation}
We complete the approximation by noting that $\hat{\bm{I}}(\tilde{\theta}^{MAP}_{m, \delta}) = \mathrm{Var}_{\tilde{\theta}^{MAP}_{m, \delta}} g(Y)$, which can be obtained e.g. by Markov Chain Monte Carlo simulation \citep{hunter2012computational}.

Putting the pieces together, Algorithm~\ref{alg:map_ergm} provides a simple procedure for performing MAP inference for pooled ERGMs under conjugate priors.  We begin by specifying the prior parameters $\bar{\tau}$ and $n_0$, and computing the mean data vector $\bar{g}(\bm{y}^{obs})$ and relative prior weight $\delta$.  The key steps are lines 3-4, which obtain the MAP estimate and associated approximate posterior variance-covariance matrix by performing the same calculations as are required for obtaining a single-graph MLE and its sample variance-covariance matrix: we simply fit to the posterior expectation $\delta\bar{\tau}+(1-\delta)\bar{g}(\bm{y}^{obs})$ instead of to an observed data value, and then adjust the information matrix to reflect the total posterior weight (prior pseudo-observations plus $m$).  Not only does this allow us to perform inference for large-$m$ data sets at no additional cost (as we did for the MLE), but it also allows us to perform Bayesian inference using algorithms and/or software implementations that were designed for maximum likelihood inference (or for first-order method-of-moments, which corresponds to maximum likelihood in this case) without additional modification.

\begin{algorithm}
\caption{MAP Inference for a Pooled ERGM Using Mean Values \label{alg:map_ergm}}
\begin{algorithmic}[1]
\REQUIRE Observed data $\bm{y}^{obs}$, prior data expectation $\bar{\tau}$ and sample size $n_0$ 
\STATE Compute $\bar{g}(\bm{y}^{obs}) =\tfrac{1}{m} \sum_{i=1}^m g(y_i)$
\STATE Let $\delta = n_0/(n_0+m)$
\STATE Find $\tilde{\theta}^{MAP}_{m, \delta} = \arg\max_\theta \left[\delta\bar{\tau}+(1-\delta)\bar{g}(\bm{y}^{obs})\right]^\intercal \theta  -  \psi( \eta(\theta) )$
\STATE Find $\widehat{\mathrm{Var}}  \tilde{\theta}^{MAP}_{m, \delta} = \tfrac{1}{m+n_0} \left[ I(\tilde{\theta}^{MAP}_{m, \delta})\right]^{-1}$
\ENSURE $\tilde{\theta}^{MAP}_{m, \delta}, \widehat{\mathrm{Var}}  \tilde{\theta}^{MAP}_{m, \delta} $ 
\end{algorithmic}
\end{algorithm}

In addition to its computational convenience, we note that the posterior expected statistic $\delta \bar{\tau} + (1 - \delta)\bar{g}(\bm{y}^{obs})$ has an intuitive geometric interpretation as a convex combination of the prior information and the observed information, with the respective weight being determined by the relative size of the prior weight $n_0$ versus $m$.  In particular, note that as $\delta\to 0$, we approach the MLE, while the prior becomes unchanged by the observed data in the limit as $\delta \to 1$.  We examine this behavior in greater detail below.  We also observe that so long as $\hat{\tau}$ lies in the relative interior of the convex hull of $\{g(y) : y \in \mathcal{Y}_n\}$, then $\tilde{\theta}^{MAP}_{m, \delta}$ exists (and is unique).  This suggests the use of conjugate MAP to address a common practical problem in ERGM inference, namely the non-existence of the MLE when the observed statistics $g(y^{obs})$ lie on the face of the convex hull of possible statistics.  In such cases, there is a direction of recession within the parameter space, with respect to which the MLE diverges; often, however, such divergent parameter values arise from very minimal information, as e.g. when a small subset of vertices in a sparse graph have no ties to each other (leading to a divergence in the corresponding homophily term).  Use of MAP inference with a small $\delta$ can improve performance in these cases by acting as a \emph{regularizer}, shrinking in extreme parameter estimates that have little support from the likelihood without otherwise greatly altering the solution.  We examine this further in Section~\ref{sec_priors}.

\subsubsection{Conjugate Prior Specification} \label{sec_priors}

Conventional research on Bayesian analysis of ERGMs focuses on priors assigned on the natural parameter space (see e.g., \citep{koskinen2004bayesian,caimo2011bayesian,koskinen.et.al:sn:2013}), whereas the ERGM conjugate prior here is actually specified in the mean-value parameter space.   This has the potential advantage that prior parameters are specified in terms of hypothetical observables (i.e., graph statistics), which are both concrete and generalizable from previously observed data; for instance, it may be easier for the analyst to specify an expected mean degree for a hypothetical network belonging to a well-studied class (e.g., friendship nominations within high schools) than to specify prior mean parameter values \emph{per se}.  By turns, given an intuition regarding plausible parameter values, it is straightforward to obtain corresponding values of $\bar{\tau}$ by simulation.  Here, we discuss some basic strategies for selecting reasonable prior parameters in practice, with the impact of prior choices being examined further in Section~\ref{sec_bayes_sim}.  

As discussed in Section~\ref{sec_map}, the specification of ERGM conjugate prior consists of two components: the \emph{a priori} expected sufficient statistics, $\bar{\tau}$, and the corresponding prior weight, $n_{0}$.  As with other exponential families, we may imagine this prior as arising from a situation in which we initially have no information regarding $\theta$ (in the sense of a limiting ``flat'' prior with $n_0 \to 0$), and then observe $n_0$ IID graph draws with mean statistics $\bar{\tau}$; our resulting state of knowledge is then summarized by the corresponding conjugate prior.  This ``prior pseudo-data'' interpretation makes the conjugate prior particularly easy to understand and communicate, and it can greatly facilitate sanity checking: for instance, if we observe that a proposed $\bar{\tau}$ value implies a mean degree far in excess of any value that could plausibly be observed in practice, then we are immediately aware of the need for refinement.

While prior specification is by nature problem specific, we here suggest several reasonable strategies for selection of $\bar{\tau}$.  Where the analyst has access to a sample of networks, $\bm{y}^{comp}$, that are similar to the network of interest (i.e., that are believed to have been produced by a similar generative process) setting $\bar{\tau}=\bar{g}(\bm{y}^{comp})$ is a natural informative choice; in this case, the posterior expectation of $g(Y)$ is shrunk towards the prior population mean.  In other cases, however, the analyst may lack such a sample, or may wish to posit a minimally informative prior that regularizes inference without strongly influencing the final estimate (a long-established tradition in Bayesian analysis, per e.g., \citep{jeffreys1961theory,hartigan1964invariant,bernardo1979reference,gelman2008weakly}, etc.).  In this context, it is useful to consider the homogeneous Bernoulli graphs (in which each edge is an IID Bernoulli trial), as a basis for the prior distribution; proposed as early as \citet{rapoport1953spread}, then later described independently by \citet{erdos1959publicationes} and \citet{gilbert1959random} as the Gilbert-Erd\H{o}s-R\'{e}nyi model in graph theoretic research, the Bernoulli graphs also arise for typical (counting measure) ERGMs as the base case where all parameters other than that associated with the edge count are equal to 0.  Given a prior expected degree $\bar{d}$ (chosen e.g. on the basis of observations of similar networks, or from prior domain knowledge), we may then set $\bar{\tau}$ by (1) drawing a sample of IID Bernoulli graphs $\bm{Y}_p^{Bern}$ with parameter $p=\bar{d}/(n-1)$, and then (2) setting the prior expectation $\bar{\tau}=\bar{g}(\bm{Y}_p^{Bern})$.  (In some cases, it may also be feasible to derive the expected statistics analytically from $p$, in which case these values may be used directly; however, exact sampling of Bernoulli graphs is extremely efficient, and a Monte Carlo approach may be easier to implement in practice.)  As the Bernoulli graphs coincide with the \emph{de facto} null model against which estimated parameters are typically assessed, setting $\bar{\tau}$ to the Bernoulli graph expectation effectively shrinks estimates towards the null model (analogously to the use of a zero-centered Gaussian or other prior in the natural parameter space), making it a reasonable default choice when more refined information is not available.

We now turn to the prior weight (``pseudo-sample size''), $n_0$.  It is convenient to consider $n_0$ via the relative prior weight, $\delta=n_0/(n_0+m)$, which quantifies the contribution of the prior to the posterior mean statistics -- the prior will dominate the data in determining the posterior when $\delta \rightarrow 1$ (i.e. $n_{0} \rightarrow \infty$), whereas a more ``objective'' analysis which lets the data ``speak for themselves'' can be obtained by letting $\delta \rightarrow 0$ (i.e. $n_{0} \rightarrow 0$, which as noted converges to the MLE). As noted above, a small-$\delta$ prior can also be viewed as a tool to regularize the model to avoid the extreme inferences resulting from data that is at or near the face of the convex hull of the sufficient statistics.  While the impact of $\delta$ on the posterior mean of the sufficient statistics is self-evident from e.g. Eq.~\ref{eq:map}, its effect in the natural parameter space is less obvious.  We examine this numerically via simulation in Section~\ref{sec_bayes_sim}.

\section{Simulation Studies} \label{sec_sim}

In this section, we conduct simulation studies to assess the behavior of the pooled MLE as $m$ becomes large, and to examine how prior specifications affect conjugate MAP inference.  To provide a realistic basis for evaluation, we base our simulated networks on Goodreau's Faux Mesa High School (FMHS) data \citep{hunter2008ergm}, a synthetic network based on proprietary data on attributions of friendship among students in a high shool in the southwestern United States \citep{resnick1997protecting}.  The FMHS network represents simulated in-school friendships among the 205 students in the school, along with their individual attributes, and was constructed to preserve the structural properties of the underlying data set.  For our study, we first fit an ERGM model to the FMSH network with the following three statistics as implemented in \citet{schmid2017exponential}: number of edges; uniform homophily by gender; and geometrically weighted edgewise shared partners (GWESP) (a common term for inducing triad closure), with the decay parameter $\lambda$ fixed at 0.25. A detailed definition of above-mentioned network statistics is in Appendix A.  Given the specified model, we first compute the MCMC MLE and treat the estimated coefficients as the networks' ``true'' parameter values $\theta_{0} = (-5.88,0.52,1.86)$; we henceforth refer to this model (i.e., ERGM distribution) as $\mathcal{M}$, from which we draw random samples for our simulation studies (i.e., $\bm{Y}\sim M$). All computations in this paper were carried out with the statistical environment R \citet{team2017r}, using the statnet libraries \citep{handcock.et.al:jss:2008,hunter2008ergm,butts2008network,butts2008social}.  \texttt{ergm} version 4.1.2 was used for all ERGM-specific computation, using default simulation and estimation settings except as otherwise noted. 

\subsection{Behavior of the MLE in Pooled-Likelihood Inference} \label{sec_sim_mle}

For our first study, we vary the sample size $m$ and examine the observed coverage rates of nominal 95\% confidence intervals for model parameters as a function of sample size. Specifically, for each value of $m$, we generate $K=1000$ datasets of size $m$ from $\mathcal{M}$, performing pooled likelihood-based inference for each sample as discussed in Section~\ref{sec_mle}.  Respective burn-in and thinning intervals of $1\times 10^6$ and $2\times 10^5$ were employed for each simulated sample (for both data simulation and MCMC-MLE inference), with MCMC-MLE termination based on the \texttt{ergm} Hotelling criterion (an autocorrelation-adjusted $T^2$ test of expected versus target statistics obtaining $p>0.5$). Table \ref{tb:coverage} presents the observed coverage rates of nominal $95\%$ confidence intervals based on the asymptotic distribution of the MLE, as estimated from the size-corrected Fisher information obtained from a pooled (single-graph) estimate.  

\begin{table}[ht]
    \centering
    \caption{Pooled MLE bias, standard error, and coverage rates of $95\%$ Wald confidence intervals as a function of sample size. \label{tb:coverage}}
    \begin{threeparttable}
    \begin{tabular}{rrrrrrrrrr}
    \hline\hline
    & \multicolumn{3}{c}{edges} & \multicolumn{3}{c}{nodematch(Gender)} & \multicolumn{3}{c}{GWESP(0.25)}\\
    \multicolumn{1}{c}{$m$}  & \multicolumn{1}{c}{Bias} & \multicolumn{1}{c}{(SE)} & \multicolumn{1}{c}{CP$^{*}$} & \multicolumn{1}{c}{Bias} & \multicolumn{1}{c}{(SE)} & \multicolumn{1}{c}{CP$^{*}$} & \multicolumn{1}{c}{Bias} & \multicolumn{1}{c}{(SE)} & \multicolumn{1}{c}{CP$^{*}$}\\
    \hline
    1 & -0.0014 & (0.138) & 0.955 & 0.0035 & (0.131) & 0.950 & -0.0097 & (0.11) & 0.949 \\ 
  2 & -0.0049 & (0.095) & 0.946 & -0.0022 & (0.087) & 0.950 & 0.0010 & (0.078) & 0.953 \\ 
  5 & -0.0001 & (0.06) & 0.955 & 0.0019 & (0.058) & 0.937 & -0.0022 & (0.048) & 0.947 \\ 
  10 & 0.0020 & (0.043) & 0.955 & -0.0012 & (0.04) & 0.951 & -0.0021 & (0.035) & 0.949 \\ 
  20 & 0.0006 & (0.031) & 0.946 & -0.0004 & (0.028) & 0.951 & -0.0007 & (0.025) & 0.935 \\ 
  30 & 0.0012 & (0.027) & 0.930 & -0.0002 & (0.023) & 0.953 & -0.0012 & (0.022) & 0.930 \\ 
  40 & 0.0005 & (0.022) & 0.942 & -0.0010 & (0.021) & 0.939 & 0.0000 & (0.018) & 0.946 \\ 
  50 & -0.0008 & (0.02) & 0.936 & 0.0000 & (0.019) & 0.932 & 0.0008 & (0.016) & 0.948 \\ 
  75 & -0.0002 & (0.017) & 0.937 & -0.0004 & (0.015) & 0.943 & 0.0005 & (0.014) & 0.943 \\ 
  100 & 0.0005 & (0.015) & 0.920 & 0.0004 & (0.014) & 0.933 & -0.0006 & (0.012) & 0.940 \\
    \hline \hline
    \end{tabular}
    \begin{tablenotes}
    \footnotesize
    \item[*] CP : Coverage Probability approximated by coverage rates of the simulated data
    \end{tablenotes}
    \end{threeparttable}
\end{table}

Table \ref{tb:coverage} shows the observed bias of the pooled MLE, as well as its standard error and the observed coverage rates of its nominal $95\%$ CIs for all three model parameters under different sample sizes ranging from 1 to 100 graphs.  Bias is negligible even for a single graph, declining to the levels of numerical noise once $m$ is greater than 5-10.  Likewise, efficiency (as measured by the standard error of the estimator) is high, and scales with $\sqrt{m}$ in the manner expected from asymptotic theory.  It is also evident that CIs based on the asymptotic distribution of Eq.~\ref{eq:ERGMasymp} perform well, maintaining approximately nominal coverage rates over a wide range of sample sizes. As a practical observation, we note that the construction of such CIs is based on statistical uncertainty, and does not take into account numerical sources of error (arising, e.g. from imperfect optimization, Monte Carlo error, etc.)\footnote{This is not strictly true, as the single-graph information matrix estimates here (produced by the \texttt{ergm} library) do incorporate some MCMC error correction.  However, it is difficult to account for all sources of numerical error, and in any event the theory of Eq.~\ref{eq:ERGMasymp} does not address it.}.  As $m\to\infty$, the statistical error becomes arbitrarily small, thus increasing the proportion of \emph{de facto} error arising from numerical approximation; put another way, it is possible to enter a regime in which our inferential precision is limited by our ability to compute the MLE (and $\hat{\bm{I}}$) rather than by the limits of our data.  To ensure accurate coverage in such extreme-$m$ scenarios, it may be necessary to adopt more stringent MCMC burn-in and thinning settings than are typically necessary for single-graph inference (where statistical uncertainty dominates), or to devise improved error estimates that better account for approximation error.  That said, we do find excellent performance for sample sizes considered here, suggesting that the problem may be limited in practice.  Further, we observe that the excellent coverage obtained for small $m$ (even $m=1$) provides practical validation of the traditional practice of using asymptotic confidence intervals in the $m=1$ case; for a review of different types of ERGM asymptotics (and their relationship to classical results) see e.g. \citet{schweinberger.et.al:ss:2020}.

\subsection{Prior Weights and MAP Inference} \label{sec_bayes_sim}

As discussed in Section~\ref{sec_priors}, choosing the relative prior weight ($\delta$) is an important aspect of the prior specification; while the choice of $\bar{\tau}$ can often be made based on either prior data or domain knowledge, the impact of $n_0$ (hence $\delta$) is less obvious.  Here, we examine the impact of $\delta$ on the MAP estimate with a particular interest in identifying prior parameter values that are likely to serve as reasonable starting points for use in regularization.  Our analysis looks first at the impact of $\delta$ on the MAP estimate itself (i.e., the extent of interpolation between the implicit prior natural parameter and the MLE), and then considers the effect of $\delta$ on the frequentist properties of the MAP estimate (bias, and the frequentist coverage of the posterior credible intervals).  

To specify a prior, we first simulate homogeneous Bernoulli random graphs on the node set of the FMHS network, given expected mean degree fixed at the average degree of all the nodes in three comparable networks (i.e. Goodreau's Faux Magnolia High School data, Faux Dixon High School data, and Faux Desert High School data \cite{hunter2008ergm}).  The observed average degree across these data sets is $1.974$, leading to an edge coefficient of $\log(\frac{p}{1-p}) = \log(\frac{\bar{d}}{n-\bar{d}-1}) = \log(\frac{1.974}{205-1.974-1}) \approx -4.63$; for the Bernoulli family, the parameters for the other two terms are set to $0$.  (We note in passing that calibration of this kind should generally be done using mean degree rather than density, as mean degree is often close to size-invariant for comparable relations while density is not; see e.g. \citet{krivitsky2011adjusting,butts:jms:2019}.) We then calculate the average network statistics of $500$ draws from this distribution, giving us the prior expected statistics $\bar{\tau} = (201.64,99.89, 3.62)$.  Since our focus is on $\delta$,  we fix our sample size at $m=1$ and vary $n_0$ to obtain the posterior inference under different values of relative prior weights.   We perform MAP estimation on $1000$ independent realizations of $Y\sim M$ for each choice of $\delta$, comparing the resulting parameter estimates to their true values ($\theta_0$) to assess the bias of MAP estimate and the frequentist coverage probability of the $95\%$ posterior credible intervals arising from the Laplace approximation to the posterior distribution.

\begin{table}[ht]
\centering
\caption{Mean MAP estimates of model parameters under different relative prior weights when $m = 1$ \label{tb:prior_weights}}
\begin{tabular}{rrrr}
  \hline\hline
 $\delta$ & edges & nodematch(Gender) & GWESP(0.25) \\
  \hline
0.0000 & -5.8866 & 0.5352 & 1.8575 \\
 0.0010 & -5.8854 & 0.5352 & 1.8560  \\
 0.0020 & -5.8836 & 0.5349 & 1.8546  \\
 0.0050 & -5.8784 & 0.5339 & 1.8501  \\
 0.0075 & -5.8736 & 0.5325 & 1.8460  \\
 0.0100 & -5.8692 & 0.5316 & 1.8423  \\
 0.0200 & -5.8519 & 0.5273 & 1.8275  \\
 0.0500 & -5.8003 & 0.5151 & 1.7830  \\
 0.1000 & -5.7181 & 0.4936 & 1.7116  \\
 0.2000 & -5.5636 & 0.4478 & 1.5763  \\
 0.3000 & -5.4213 & 0.3994 & 1.4475  \\
 0.4000 & -5.2891 & 0.3485 & 1.3225  \\
 0.5000 & -5.1640 & 0.2941 & 1.1969  \\
 0.7500 & -4.8787 & 0.1471 & 0.8424  \\
 0.9000 & -4.7232 & 0.0531 & 0.5206  \\
 0.9500 & -4.6739 & 0.0211 & 0.3394  \\
 0.9900 & -4.6352 & -0.0044 & 0.0895  \\
 1.0000 & -4.6258 & -0.0108 & -0.0139 \\ 
  \hline\hline
\end{tabular}
\end{table}

\begin{figure}
\centering
\includegraphics[width=5.5in]{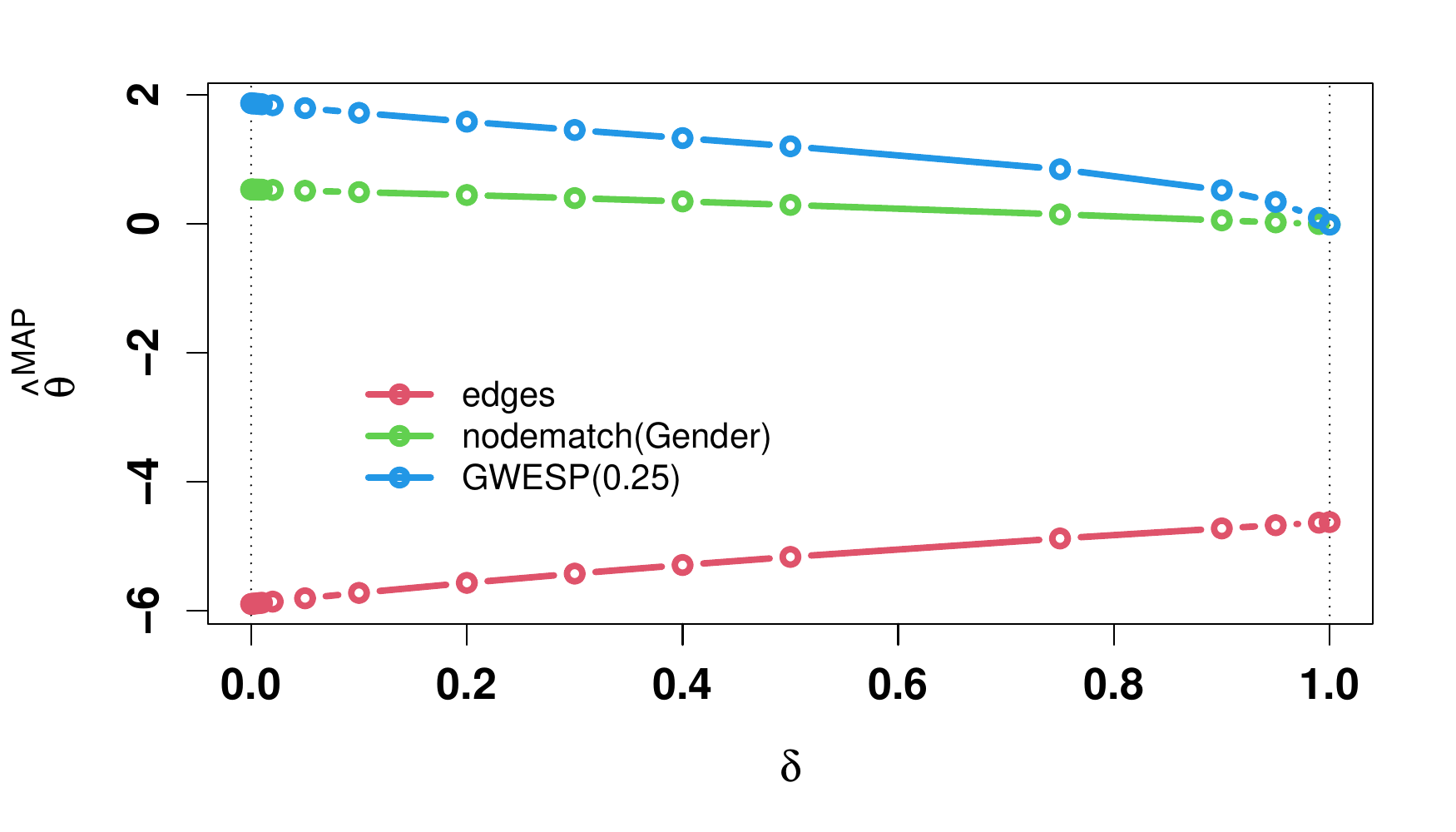}
\caption{Average MAP estimates for $Y\sim M$, by $\delta$. Solutions interpolate between the natural parameters corresponding to the prior ($\delta=1$) and the MLE ($\delta=0$); shrinkage is nearly linear in $\delta$ near the non-informative limit.   \label{pic:map_meanest}}
\end{figure}

We begin by examining the impact of $\delta$ on the MAP estimate.  As noted above, the MAP estimate must interpolate between the natural parameter equivalent of $\bar{\tau}$ at $\delta=1$ and the MLE at $\delta=0$; equivalently, we may think of the conjugate prior as shrinking the estimate towards the (natural parameter equivalent) of the prior expectation.  The detailed pattern of shrinkage is depicted in fig.~\ref{pic:map_meanest}, which shows that parameters change roughly linearly over most of the unit interval, with the most extreme changes occurring near $\delta=1$.  Importantly, shrinkage is approximately linear near the non-informative limit ($\delta=0$), suggesting that small differences in choice of $\delta$ do not have a large impact on the posterior mode (a convenient property when selecting minimally informative priors for regularization purposes).  A more quantitative picture emerges from Table~\ref{tb:prior_weights}, which shows the mean MAP estimates for each parameter as a function of $\delta$.  We observe that choosing $\delta \leqslant 0.02$ yields estimates that are extremely close to the MLE (agreeing to 2-3 decimal places), while still placing sufficient weight on the prior to be useful for regularization (i.e., to ensure that the mean value parameter lies in the relative interior of the convex hull of possible statistics).

\begin{table}[ht]
    \centering
    \caption{Frequentist properties of MAP estimate and Laplace approximation for model parameters under different relative prior weights with sample size fixed at $m = 1$ \label{tb:prior_weights_bias_coverage}}
    \begin{threeparttable}
    \begin{tabular}{rrrrrrrr}
    \hline
    & & \multicolumn{2}{c}{edges} & \multicolumn{2}{c}{nodematch(Gender)} & \multicolumn{2}{c}{GWESP(0.25)}\\
    \multicolumn{1}{c}{$\delta$} & \multicolumn{1}{c}{$n_0^{\dag}$} & \multicolumn{1}{c}{Bias} & \multicolumn{1}{c}{(CP$^{\ddag}$)} & \multicolumn{1}{c}{Bias} & \multicolumn{1}{c}{(CP$^{\ddag}$)} & \multicolumn{1}{c}{Bias} & \multicolumn{1}{c}{(CP$^{\ddag}$)}\\
    \hline
 0.0000 & 0.000 & -0.0014 & 0.955 & 0.0035 & 0.950 & -0.0097 & 0.949  \\
 0.0010 & 0.001 & -0.0003 & 0.953 & 0.0035 & 0.949 & -0.0111 & 0.948  \\
 0.0020 & 0.002 & 0.0015 & 0.953 & 0.0032 & 0.952 & -0.0126 & 0.949  \\
 0.0050 & 0.005 & 0.0067 & 0.953 & 0.0022 & 0.952 & -0.0171 & 0.947  \\
 0.0075 & 0.008 & 0.0115 & 0.956 & 0.0008 & 0.952 & -0.0212 & 0.942  \\
 0.0100 & 0.010 & 0.0159 & 0.956 & -0.0001 & 0.949 & -0.0249 & 0.945  \\
 0.0200 & 0.020 & 0.0333 & 0.946 & -0.0044 & 0.948 & -0.0397 & 0.930  \\
 0.0500 & 0.053 & 0.0848 & 0.900 & -0.0166 & 0.952 & -0.0842 & 0.894  \\
 0.1000 & 0.111 & 0.1670 & 0.733 & -0.0381 & 0.942 & -0.1555 & 0.654  \\
 0.2000 & 0.250 & 0.3215 & 0.157 & -0.0839 & 0.911 & -0.2909 & 0.098  \\
 0.3000 & 0.429 & 0.4638 & 0.001 & -0.1323 & 0.818 & -0.4196 & 0.000  \\
 0.4000 & 0.667 & 0.5961 & 0.000 & -0.1832 & 0.604 & -0.5447 & 0.000  \\
 0.5000 & 1.000 & 0.7211 & 0.000 & -0.2376 & 0.269 & -0.6703 & 0.000  \\
 0.7500 & 3.000 & 1.0064 & 0.000 & -0.3846 & 0.000 & -1.0248 & 0.000  \\
 0.9000 & 9.000 & 1.1619 & 0.000 & -0.4786 & 0.000 & -1.3466 & 0.000  \\
 0.9500 & 19.000 & 1.2112 & 0.000 & -0.5106 & 0.000 & -1.5278 & 0.000  \\
 0.9900 & 99.000 & 1.2500 & 0.000 & -0.5361 & 0.000 & -1.7776 & 0.000  \\
 1.0000 &  $\infty$ & 1.2593 & 0.000 & -0.5425 & 0.000 & -1.8811 & 0.000  \\
   \hline\hline
    \end{tabular}
    \begin{tablenotes}
    \scriptsize{
    \item[$\dag$] CP: Coverage Probability approximated by coverage rates of the simulated data; Laplace approximation is not applicable to $\delta = 1$ because in that case the prior dictates the posterior inference, the posterior interval degenerates to a point mass 
    \item[$\ddag$] $n_{0}$ is the equivalent sample size contained in the prior given its relative weight $\delta$}
    \end{tablenotes}
    \end{threeparttable}
\end{table}

We now turn to the frequentist properties of the MAP estimate, as a function of $\delta$.  Here we compare the MAP estimate (and the 95\% posterior intervals arising from the Laplace approximation) to the coefficients of true model $\mathcal{M}$, which is $\theta_{0} = (-5.89,0.53,1.87)$; at the other extreme, we have the natural parameter equivalent of the location of the conjugate prior ($(-4.63, 0, 0)$). Table \ref{tb:prior_weights_bias_coverage} shows the estimated bias and frequentist coverage probability for our simulation sample, as a function of $\delta$.  As can be seen, bias is minimal until $\delta \approx 0.02$, becoming substantial for $\delta > 0.1$.  Likewise, the 95\% posterior intervals maintain good frequentist calibration until roughly $\delta \approx 0.02$, though coverage degrades rapidly thereafter.  For regularizing/minimally informative applications, a choice of $n_0 \approx 0.01$ (giving the prior approximately 1\% of the weight of a single graph observation) would seem to be a reasonable starting point.

\section{Applications} \label{sec_apps}

To demonstrate the pooled ERGM/conjugate prior analysis in practice, we provide two illustrative applications.  The first is to the analysis of brain functional connectivity networks, where we seek a common model for brain structure across individuals.  The second considers the use of ERGMs to model variation in protein structures obtained by X-ray crystallography, in this case using hen egg-white lysozyme (a widely studied reference protein).  In each case, we show how the approach used here facilities the simultaneous analysis of multiple networks, and provides a fast and simple means of performing Bayesian inference.

\subsection{Pooled ERGM Analysis for Brain Functional Connectivity Networks} \label{sec_brain}

The study of group-based brain functional connectivity networks has become a topic of increasing interest in neuroscience, due the need to characterize both central tendencies and patterns of variation in interactions among brain regions. Importantly, it is of interest not only to measure specific or mean interactions, but to be able to characterize the distributions of interaction patterns arising under particular conditions, and/or within particular subpopulations.  ERGMs have been identified as a promising tool for this purpose, due to their ability to assess how local brain network features give rise to the global structure, and due to their capacity to account for both heterogeneity and dependence among interactions \citep{simpson2011exponential, simpson2013analyzing}. 

Brain functional connectivity networks often exhibit both functional segregation and integration \citep{rubinov2010complex}, where functional segregation in the brain is the ability for specialized processing to occur within densely interconnected groups of brain regions, while functional integration corresponds to the ability to rapidly amalgamate specialized information from scattered brain regions. As an attempt to produce a model with appropriate network sufficient statistics that are able to capture those two concurrent opposing driving forces, \citet{simpson2012exponential} proposed to first select the ``best'' metrics from a broader set of potential candidates identified in the literature using model selection techniques for ERGMs, then refit the networks of all subjects with those ``best' metrics.  They then employed the mean (respectively median) of the resulting individual estimates as estimates of a global, group-level ``representative'' whole-brain connectivity network model (which they refer to as a ``mean'' (respectively) ``median'' ERGM).  This method of amalgamating models in the natural parameter space is straightforward and intuitive, but has several disadvantages: as shown in Eq.~\ref{eq:mleopt}, the appropriate pooling for a joint ERGM occurs in the mean value parameter space rather than the natural parameter space; separate estimation of an ERGM for each individual is computationally expensive (and, for the MLE, may encounter problems if some individuals' networks have statistics that lie on the face of the convex hull of potential statistics); the statistical properties of the amalgamated model are unclear (especially in the median case); and model selection by this approach does not exploit the joint likelihood (which may lead to an inferior pooled model).

By contrast, a pooled-ERGM approach provides a more principled and computationally efficient alternative to the mean/median ERGM approach.  For large $n$, the properties of the pooled estimates and their confidence intervals are ensured by the large sample theory of exponential families, and as shown in Section~\ref{sec_sim_mle} good results can be obtained with even modest numbers of graphs.  Moreover, instead of having to fit each observed network separately, as proposed in \citet{simpson2012exponential} (with the risk that the MLE will not exist in one or more cases), exactly one ERGM fit is required (and the target statistics for that fit lie on the face of the convex hull only if all input networks do as well).  Moreover, the ability to use conjugate-MAP inference for pooled ERGMs provides an inexpensive way of obtaining approximate Bayesian answers where desired, or (when viewing the prior as a regularizer) obtain regularized likelihood estimates.  Here, we demonstrate all three approaches in the context of brain functional connectivity networks, building on prior work by \citet{simpson2011exponential,simpson2012exponential}.  Due to the large number of model fits required for cross-validation, we use \texttt{ergm}'s stochastic approximation method for estimation in this section, with all Markov chains having a thinning interval of $5 \times 10^4$ following $2 \times 10^5$ burn-in iterations.

\subsubsection{Data}
\begin{figure}
\centering
\includegraphics[width=0.75\textwidth]{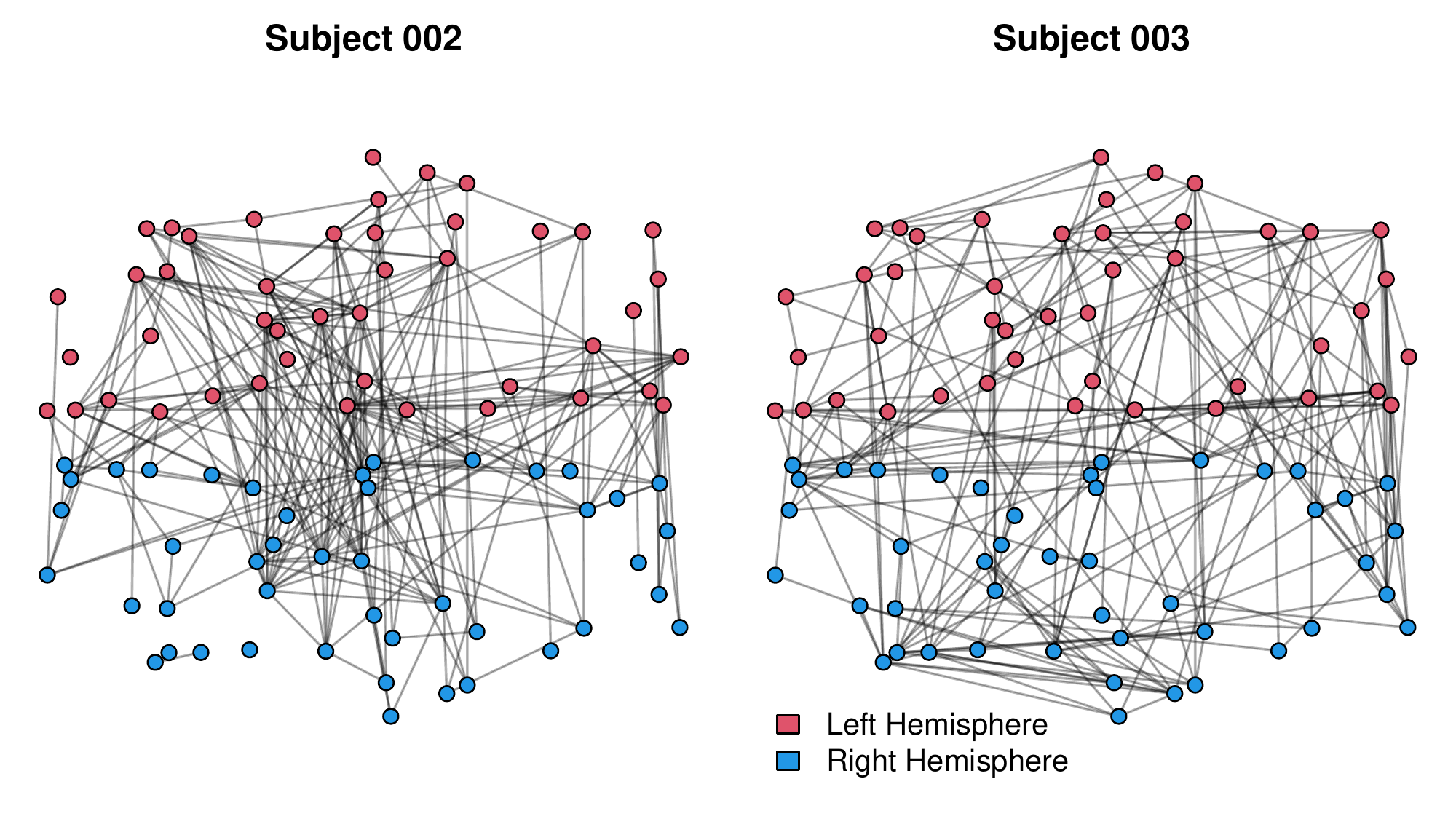}
\caption{Brain connectivity networks of Subjects 002 and 003. Colors (red, blue) indicate the different hemispheres (left, right); node coordinates are based on an non-metric MDS \citep{venables.ripley:bk:2002} solution for distances between regions of interest.   \label{pic:brain_002_003}}
\end{figure}

We consider the data reported in \citet{simpson2011exponential,simpson2012exponential}, which includes brain functional connectivity networks among 10 normal subjects (5 female, average age: 27.7 years old, standard deviation: 4.7 years) who were part (Subject No. 002, 003, 005, 008, 009, 010, 012, 013, 016, 021) of a larger functional MRI study of age-related changes in cross-modal deactivations \citep{peiffer2009aging}. Fig.~\ref{pic:brain_002_003} depicts the brain connectivity networks of subjects 002 and 003, illustrating both common properties (e.g., clustering, increased probability of ties within brain regions) and heterogeneity across networks; here, we are interested in capturing this distribution via an ERGM form.   Note that brain connectivity networks are defined on equivalent sets of nodes, which here correspond to 90 prespecified brain regions (ROIs -- Regions of Interest), according to the Automated Anatomical Labeling atlas (AAL) \citep{tzourio2002automated}. Each of these 10 brain connectivity networks is represented by binary adjacency matrix, in which element $(i,j)$ denotes the presence or absence of a functional connection between node $i$ and node $j$. The establishment of binary functional connections was done by thresholding the temporal correlation coefficient adjusted for motion and physiological noise (see \citep{hayasaka2010comparison, simpson2011exponential} for further details), and hence those brain networks are undirected by construction. The thresholds were selected by the original authors at the subject level to make each network with $\frac{\log(n)}{\log(\bar{d})} \approx 2.8$, or equivalently $\bar{d} \approx n^{1/2.8}$, where $n$ is the total number of nodes (here, $n=90$). 

Covariate information associated with these networks includes not only the nodal covariates \emph{Hemisphere} and \emph{Area}, but also an edge-level covariate for the spatial distance matrix among the ROIs (Mean: 76.28 mm, SD: 28.93 mm). The 90 regions are divided symmetrically across left and right hemispheres, with each hemisphere consisting of 28 areas as presented in Table \ref{tb:brain_area}. 

\begin{table}[H]
\centering
\caption{Number of ROIs in each area of the brain; names follow Simpson et al. (2011) \label{tb:brain_area}}
\begin{tabular}{lrr}
  \hline

Area & Left Hemisphere & Right Hemisphere \\ 
  \hline
Amygdala & 1 & 1 \\ 
  Angular & 1 & 1 \\ 
  Calcarine & 1 & 1 \\ 
  Caudate & 1 & 1 \\ 
  Cuneus & 1 & 1 \\ 
  Fusiform & 1 & 1 \\ 
  Heschl & 1 & 1 \\ 
  Hippocampus & 1 & 1 \\ 
  Insula & 1 & 1 \\ 
  Lingual & 1 & 1 \\ 
  Olfactory & 1 & 1 \\ 
  Pallidum & 1 & 1 \\ 
  Paracentral & 1 & 1 \\ 
  ParaHippocampal & 1 & 1 \\ 
  Postcentral & 1 & 1 \\ 
  Precentral & 1 & 1 \\ 
  Precuneus & 1 & 1 \\ 
  Putamen & 1 & 1 \\ 
  Rectus & 1 & 1 \\ 
  Rolandic & 1 & 1 \\ 
  Supp & 1 & 1 \\ 
  SupraMarginal & 1 & 1 \\ 
  Thalamus & 1 & 1 \\ 
  Parietal & 2 & 2 \\ 
  Cingulum & 3 & 3 \\ 
  Occipital & 3 & 3 \\ 
  Temporal & 5 & 5 \\ 
  Frontal & 9 & 9 \\ 
   \hline
\end{tabular}
\end{table}

\subsubsection{Model specification}
Connectedness, local clustering and global efficiency were introduced as the key components in previous work on brain connectivity network modeling \citep{simpson2011exponential, simpson2012exponential}, with the latter two being proposed as proxies for functional segregation and functional integration respectively. As such, their joint effects are modeled explicitly as a combination of network statistics: edge count (\texttt{Edges}), GWESP, and \emph{geometrically weighted null shared partners} (GWNSP). Such a model specification yields a homogeneous ERGM that is permutation invariant \citep{frank1986markov, schweinberger.et.al:ss:2020}, which leaves covariate information underutilized, and in turn makes the estimation difficult and unstable due to multimodality of the distribution \citep{snijders2002markov}. Similar to the multicollinearity issue in regression, it can be problematic to include two closely correlated network statistics in an ERGM model, and the presence of GWESP and GWNSP in previous models is found to be associated with convergence issue in the present case.  Here, we thus modify and extend the homogeneous model used in prior work by incorporating node-level heterogeneity and distance effects associated with the spatial structure of the brain, along with a less collinear combination of GWESP and graphletCount(1) terms to capture dependence. Specifically, we include as covariates: a homophily effect for hemisphere (\texttt{hemisphere-nodematch}), as introduced in \citet{sinke2016bayesian}; a mixing effect for brain regions (\texttt{Area-nodemix}) as a measure of the strength of interaction between the brain regions belonging to different areas of the brain; and a dyadic covariate that controls for spatial proximity (\texttt{log.spatial.dist-edgecov}), implemented by an effect for the log of the distance between regions (a common choice for modeling geographical effects, e.g., \citep{daraganova2012networks}).  In addition to providing substantive insight into the drivers of connectivity, we also observe that such covariate effects also improve model performance by separating clustering and bridging due to physical brain structure from emergent network properties arising from dependence effects. The latter are captured by two effects.  First, a GWESP term with decay fixed at $0.5$ (\texttt{GWESP, $\phi = 0.5$}) is included to capture residual tendencies towards endogenous local clustering net of controls, and second, a \texttt{graphletCount(1)} term \citep{nebil2015ergm} helps capture open two-path structure (aka \emph{local bridging}) like that previously examined using GWNSP in the models of \citet{simpson2011exponential, simpson2012exponential}. 

\subsubsection{Results}

Table \ref{tb:brain_fit_mean_stats_coef} presents maximum likelihood estimates of model coefficients and associated standard errors for the group-based brain connectivity network model under pooling, enabling us to infer the extent to which each of the proposed effects shapes the overall distribution of networks across test subjects.  We see a positive and statistically significant parameter estimate for the GWESP statistic, indicating high levels triadic closure net of spatial and anatomical features; this is compatible with the theory of functional segregation proposed in prior work.  Likewise, we see that bridging is significantly disfavored (i.e., a negative effect for graphlet 1), suggesting that open triads tend not persist (net of other factors).  In estimating mixing effects, we aggregate all areas other than \emph{Frontal} and \emph{Temporal} to a single level as ``Others'' due to the small sizes of these regions, providing a tripartite mixing structure; we see inhibition of ties between different areas, and null or positive tendencies towards formation of within-area ties, which provides additional evidence for functional segregation.  We note that these effects persist net of the overall inhibition of ties between more distant regions, with tie probability declining (\emph{ceteris paribus}) as approximately one over the inverse of the distance between nodes.  An important exception is the case of cross-hemispheric interactions, which are actually favored (the negative nodematch indicating that within-hemispheric interactions are disfavored relative to those that cross hemispheres).  This can be viewed as an indicator of functional integration, with the need for coordination across hemispheres working against the general tendency against long-range ties.  Care is required in the quantitative interpretation of the positive \texttt{Edges} coefficient, given the existence of \texttt{log.spatial.dist-edgecov}. Specifically, note that mean of pairwise distances of the ROIs is 76.28 mm and hence at the mean $\log(76.28) \approx 4.334 $, we have $\Pr(Y_{i,j} = 1 | Y_{i,j}^{c} = y_{i,j}^{c} ) = \frac{\exp (1.193-1.081\times 4.334) }{1 + \exp (1.193-1.081\times 4.334)} \approx 0.029$, conditional on the rest of the graph and all other effects held at zero, meaning that the baseline conditional probability of observing an edge (not involved in the creation of other network statistics included in the model) between pairs of regions at the average distance is still very low, as expected for sparse graphs.   

\begin{table}
\begin{center}
 \caption{Model parameter estimates and standard errors for pooled-ERGM analysis of brain functional connectivity networks \label{tb:brain_fit_mean_stats_coef}}
\begin{tabular}{lrrl}
\hline\hline
 Term & Estimate & (s.e.) &  \\
\hline
\texttt{Edges}                     &  1.193 & (0.180) & $^{****}$  \\
\texttt{GWESP, $\phi = 0.5$}       &  1.622 & (0.044) & $^{****}$   \\
\texttt{graphletCount(1)}          & -0.182 & (0.008) & $^{****}$   \\
\texttt{log.spatial.dist-edgecov}  & -1.081 & (0.032) & $^{****}$   \\
\texttt{hemisphere-nodematch}      & -0.337 & (0.043) & $^{****}$   \\
\texttt{Frontal.Others-nodemix}    & -0.270 & (0.055) & $^{****}$   \\
\texttt{Others.Others-nodemix}     & -0.061 & (0.057) &     \\
\texttt{Frontal.Temporal-nodemix}  & -0.277 & (0.106) & $^{**}$   \\
\texttt{Others.Temporal-nodemix}   & -0.351 & (0.074) & $^{****}$   \\
\texttt{Temporal.Temporal-nodemix} &  0.457 & (0.101) & $^{****}$   \\
\hline
\multicolumn{4}{c}{\scriptsize{$^{****} \: p<0.0001$, $^{***} \: p<0.001$, $^{**} \: p<0.01$, $^{*} \: p<0.05$}}\\
\hline \hline
\end{tabular}
\end{center}
\end{table}

\subsection{Approximate Conjugate Bayesian Analysis of Brain Functional Connectivity Networks}

In this subsection, we demonstrate how one can conduct approximate conjugate Bayesian analysis as introduced in Section~\ref{sec_map} for the dual purposes of approximating full Bayesian analysis and regularization. The construction of the prior is crucial regardless of the ultimate purpose. We adopt the simulation-based approach of Section~\ref{sec_priors} to specify the prior by noting that $\bar{d} \approx n^{1/2.8}$ by construction (i.e., choice of correlation threshold) for all brain functional connectivity networks in this dataset, and thus we set $\bar{\tau}$ is set to be equal to the mean of network sufficient statistics under a Bernoulli graph with $p = \frac{\bar{d}}{n-1} \approx \frac{n^{1/2.8}}{n-1} \approx 0.056$ ($n$ = 90). The selection of relative weight $\delta$ is subject to vary depending upon the purpose, which is explored and discussed in detail with examples. 

\subsubsection{MAP Estimation for the Pooled Model}
In the absence of strong \emph{a priori} information regarding almost all aspects of the brain functional connectivity networks except for the mean degree, it is advisable to incorporate weak prior information; we do this by assigning a small value to the hyper-parameter $\delta$, in this case setting $\delta=0.02$. Given a specified prior, we conduct Bayesian inference based on Algorithm~\ref{alg:map_ergm}, the resulting parameters being shown in Table \ref{tb:brain_fit_bayesian}. 

\begin{table}
\begin{center}
 \caption{MAP estimates and posterior standard deviations, conjugate Bayesian analysis of brain functional connectivity networks. \label{tb:brain_fit_bayesian}}
\begin{tabular}{lrrc}
\hline\hline
Term & Estimate & (s.d.) &  $95\%$ credible interval \\
\hline
\texttt{Edges}                      &  1.277 & (0.196) & (0.893,1.662) \\
\texttt{GWESP, $\phi = 0.5$}        &  1.584 & (0.042) & (1.502,1.665) \\
\texttt{graphletCount(1)}           & -0.186 & (0.008) & (-0.202,-0.170) \\
\texttt{log.spatial.dist-edgecov}   & -1.070 & (0.036) & (-1.141,-0.999) \\
\texttt{hemisphere-nodematch}       & -0.334 & (0.048) & (-0.427,-0.241) \\
\texttt{Frontal.Others-nodemix}     & -0.326 & (0.078) & (-0.478,-0.174) \\
\texttt{Others.Others-nodemix}      & -0.112 & (0.073) & (-0.255,0.032) \\
\texttt{Frontal.Temporal-nodemix}   & -0.287 & (0.126) & (-0.533,-0.040) \\
\texttt{Others.Temporal-nodemix}    & -0.390 & (0.087) & (-0.560,-0.220) \\
\texttt{Temporal.Temporal-nodemix}  &  0.466 & (0.120) & (0.231,0.700) \\
\hline\hline
\end{tabular}
\end{center}
\end{table}

The parameter estimates from the Bayesian analysis are very similar to those of the pooled MLE, supporting the same qualitative conclusions.  However, imposing a prior on the parameter vector permits interpretation of the results in terms of Bayesian answers, which may be useful in some settings; we may also use the Laplace approximation to sample from the approximate posterior, enabling us to obtain e.g. posterior predictive distributions for network properties that take into account uncertainty in the model parameters.  

\subsubsection{Regularizing ERGMs with MAP}

As noted above, the MLE for the natural parameter of an exponential family distribution does not exist when the observed sufficient statistics lie on the relative boundary of $C$, the convex hull of the set of possible values of sufficient statistics.  A common case of this type in ERGM modeling arises when mixing or differential nodematch parameters are specified for networks containing many small subgroups; if any of the associated statistics are equal to 0 (e.g., there are no observed ties between two groups), then the likelihood has no finite maximizer with respect to the respective directions in the natural parameter space.  In the context of the brain connectivity networks, we observe that there are many small areas containing few nodes, potentially leading to such a circumstance.   For instance, consider an extension of our previous model intended to quantify the mixing pattern between nodes in the  \emph{Occipital} and \emph{Cingulum} areas; we may do so by augmenting $\mathcal{M}_{1}$ with \texttt{nodemix} terms involving Occipital and Cingulum, with all other terms in the model unchanged. We denote this model as $\mathcal{M}_{2}$.  It happens, however, that there there are no edges observed between Occipital and Cingulum for any of the networks in the dataset, and hence the vector of mean observed network sufficient statistics is no longer located in $rint(C)$ (as the \texttt{Occipital.Cingulum-nodemix} value of $0$ is smallest possible value that can be obtained).  From an optimization perspective, we are unable to obtain a finite estimate for model coefficients of this augmented model, because the likelihood can always be further optimized by letting the vector of candidate estimates of model coefficients move towards the direction of recession.  Statistically, this reflects the non-existence of the MLE.  We now show such issues can be resolved by incorporating an appropriate conjugate prior into the inference to regularize the model and thus avoid extreme inferences on model parameters.

We construct a conjugate prior in the form of \eqref{eq:ergm_conjugate_prior}, where hyper-parameter $\bar{\tau}$ is determined by calculating the mean of network sufficient statistics observed on $1000$ independent random realizations of Bernoulli random graphs with $p = 0.056$.  As our goal here is regularization, we view the prior as a convenient penalty function (rather than as a formal statement of prior knowledge), and treat $\delta$ as a hyperparameter subject to optimization.  Given our pooled setting, it is natural to evaluate model performance by cross-validation (CV); specifically, we vary $\delta$ (or, equivalently, the prior sample size $n_0$), computing the expected squared Hamming error for each graph under leave-one-out CV based on 1000 draws from each simulated model, and select the value that minimizes the expected loss on the held-out networks.  

The results of the hyperparameter tuning process are shown in Table~\ref{tb:cverr}.  As expected, the unregularized MLE ($n_0=\delta=0$) yields suboptimal performance, with improvements obtained until $n_0=0.004$ ($\delta=0.0004$).  Further increases in prior weight (here interpreted as the strength of the penalty function) result in diminished performance, as the fitted model is drawn towards the prior mean.  We thus select $\delta=0.0004$ for subsequent analysis.  

\begin{table}
\begin{center}
\caption{Leave-one-out cross validation error for regularized inference on $\mathcal{M}_2$ as a function of $n_0$.}
\label{tb:cverr}
\begin{tabular}{rrr}
\hline\hline
$n_0$ & $\delta$ & CV Error \\ \hline
0.000 & 0.0000 & 612255.803 \\
0.001 & 0.0001 & 616388.530 \\
0.002 & 0.0002 & 611733.729 \\
0.004 & 0.0004 & 610657.580 \\
0.008 & 0.0009 & 617267.178 \\
0.016 & 0.0017 & 616974.051 \\
0.031 & 0.0035 & 620343.917 \\
0.062 & 0.0069 & 610740.156 \\
0.125 & 0.0137 & 612219.087 \\
0.250 & 0.0270 & 614275.772 \\
0.500 & 0.0526 & 618156.374 \\
1.000 & 0.1000 & 625175.603 \\
2.000 & 0.1818 & 636375.903 \\
4.000 & 0.3077 & 650507.803 \\
8.000 & 0.4706 & 670991.412 \\
\hline\hline
\end{tabular}
\end{center}
\end{table}

We may now perform penalized maximum likelihood inference, using the tuned conjugate prior as a regularizer.  Table~\ref{tb:reg_no_reg_comparison} shows the corresponding parameter estimates, standard errors, and significance levels for model $\mathcal{M}_2$.  As expected, the results for the shared effects (triangulation, spatial interaction, bridging, and hemispheric interaction) after breaking out the additional brain areas remain very similar to what was seen in the unregularized MLE for the collapsed model, though we now have a more complete description of the mixing pattern among localized areas.  Importantly, we also observe that the Occipital/Cingulum parameter (for which the MLE does not exist) is now well-characterized.  As we would expect from the fact that none of the observed networks had Occipital/Cingulum ties, the estimated coefficient is significantly negative; however, the magnitude is now plausible (and in line with the other observed effects).

\begin{table}
\begin{center}
\caption{Model parameter estimates and standard errors for regularized inference on $\mathcal{M}_2$.}
\label{tb:reg_no_reg_comparison}
\begin{tabular}{lrrl}
\hline \hline
Term &  Estimate & (s.e.) &    \\
\hline
\texttt{Edges}                      &  1.986 & (0.224) & $^{****}$   \\
\texttt{GWESP, $\phi = 0.5$}        &  1.623 & (0.045) & $^{****}$   \\
\texttt{graphletCount(1)}           & -0.166 & (0.007) & $^{****}$   \\
\texttt{log.spatial.dist-edgecov}   & -1.051 & (0.036) & $^{****}$   \\
\texttt{hemisphere-nodematch}       & -0.328 & (0.044) & $^{****}$   \\
\texttt{Cingulum.Frontal-nodemix}   & -1.109 & (0.142) & $^{****}$   \\
\texttt{Frontal.Frontal-nodemix}    & -1.006 & (0.128) & $^{****}$   \\
\texttt{Cingulum.Occipital-nodemix} & -2.311 & (0.351) & $^{****}$   \\
\texttt{Frontal.Occipital-nodemix}  & -0.638 & (0.189) & $^{***}$   \\
\texttt{Occipital.Occipital-nodemix}& -0.135 & (0.153) &     \\
\texttt{Cingulum.Others-nodemix}    & -1.310 & (0.130) & $^{****}$   \\
\texttt{Frontal.Others-nodemix}     & -1.370 & (0.129) & $^{****}$   \\
\texttt{Occipital.Others-nodemix}   & -1.290 & (0.136) & $^{****}$   \\
\texttt{Others.Others-nodemix}      & -1.069 & (0.124) & $^{****}$   \\
\texttt{Cingulum.Temporal-nodemix}  & -1.325 & (0.266) & $^{****}$   \\
\texttt{Frontal.Temporal-nodemix}   & -1.278 & (0.153) & $^{****}$   \\
\texttt{Occipital.Temporal-nodemix} & -1.767 & (0.374) & $^{****}$   \\
\texttt{Others.Temporal-nodemix}    & -1.340 & (0.131) & $^{****}$   \\
\texttt{Temporal.Temporal-nodemix}  & -0.514 & (0.148) & $^{***}$   \\
\hline
\multicolumn{4}{c}{\scriptsize{$^{****}p<0.0001$, $^{***}p<0.001$, $^{**}p<0.01$, $^*p<0.05$}}\\
\hline\hline
\end{tabular}
\end{center}
\end{table}

\subsection{Analysis of Lysozyme Structure Networks via Pooled ERGMs} \label{sec_protein}

The functions of proteins and other macromolecules are heavily influenced by their three-dimensional structure.  With the increasing sophistication of both experimental technique and molecular modeling, new methods for analyzing the growing body of protein structure data are of increasing interest.  Network analytic methods have emerged as particularly useful tools for this purpose, providing a rich representation for topological complexity while still offering substantial coarsening relative to atomistic structure.  Among other applications, network representations of protein structure have been used to identify functionally important residues \citep{amitai2004network}, summarize protein dynamics \citep{bode2007network}, identify functionally significant sub-units \citep{chakrabarty2016naps}, distinguish active site conformations \citep{cross.et.al:bio:2020}, and characterize structural differences between protein families \citep{unhelkar.et.al:bba:2017}.

One potential application of ERGMs in the context of protein structure is the characterization of  variation within structures of the same protein (either in equilibrium, or in different functional or measurement contexts).  ERGMs were first applied to protein structure networks by \citet{nebil2015ergm}, who used them to control for intrinsic molecular features (e.g., chain membership) while testing hypotheses regarding fold-specific structure.  In more recent work,  ERGMs have been employed to characterize transient structure in intrinsically disordered proteins \citep{grazioli.et.al:fmb:2019}, and to model protein aggregation \citep{grazioli.et.al:jpcB:2019,yu.et.al:nsr:2020}.  Here, we consider the problem of characterizing variation in measured protein structures obtained via X-ray crystallography (the primary workhorse technique of modern structural biology).  While it is common to treat globular proteins as having a native fold associated with a single three-dimensional structure obtained via crystallographic methods (or, more rarely, Nuclear Magnetic Resonance, neutron scattering, or cryo-EM), proteins in solution are extremely dynamic; even in a crystallographic context, repeated crystallization of the same protein will often yield slightly different structures.\footnote{In fact, the same crystal frequently contains several distinct conformations within a single asymmetric unit.}  Currently, this variation is not well-characterized, and is often ignored (with a single conformation selected as ``the'' structure of the protein).  Statistically, it is natural to think of these observed structures as being drawn from a broader distribution of low-energy conformations, and to attempt to model this distribution using the measured conformations.  

\begin{figure}
\includegraphics[]{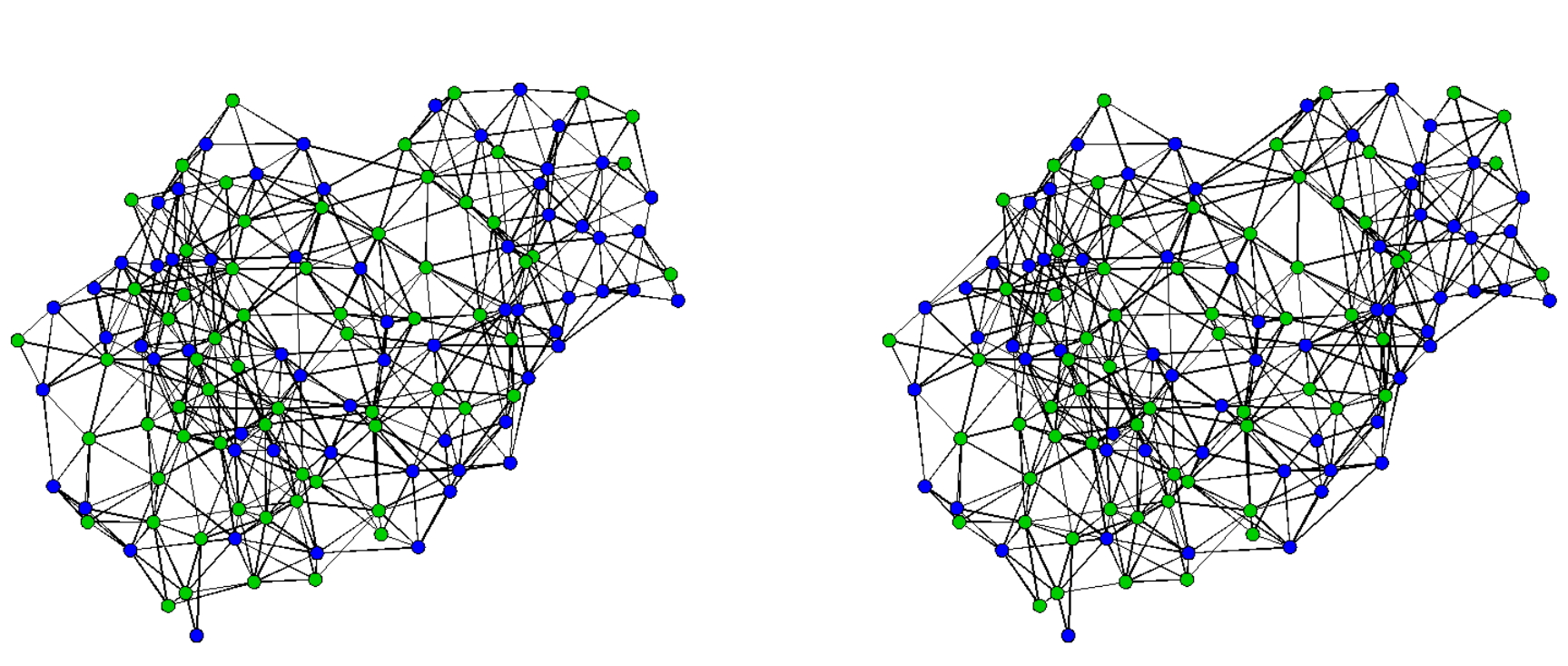}
\caption{2D representation of Lysozyme structure networks (PDB 1AKI \citep{Artymiuk:a20987}; Left), (PDB 1BHZ \citep{Ramin:ha0170}; Right). Colors distinguish nonpolar (green) versus polar (blue) residues; node coordinates determined via topology and are not based on physical position. \label{pic:lysozyme}}
\end{figure}

\begin{figure}
\centering
\includegraphics[width=5.5in]{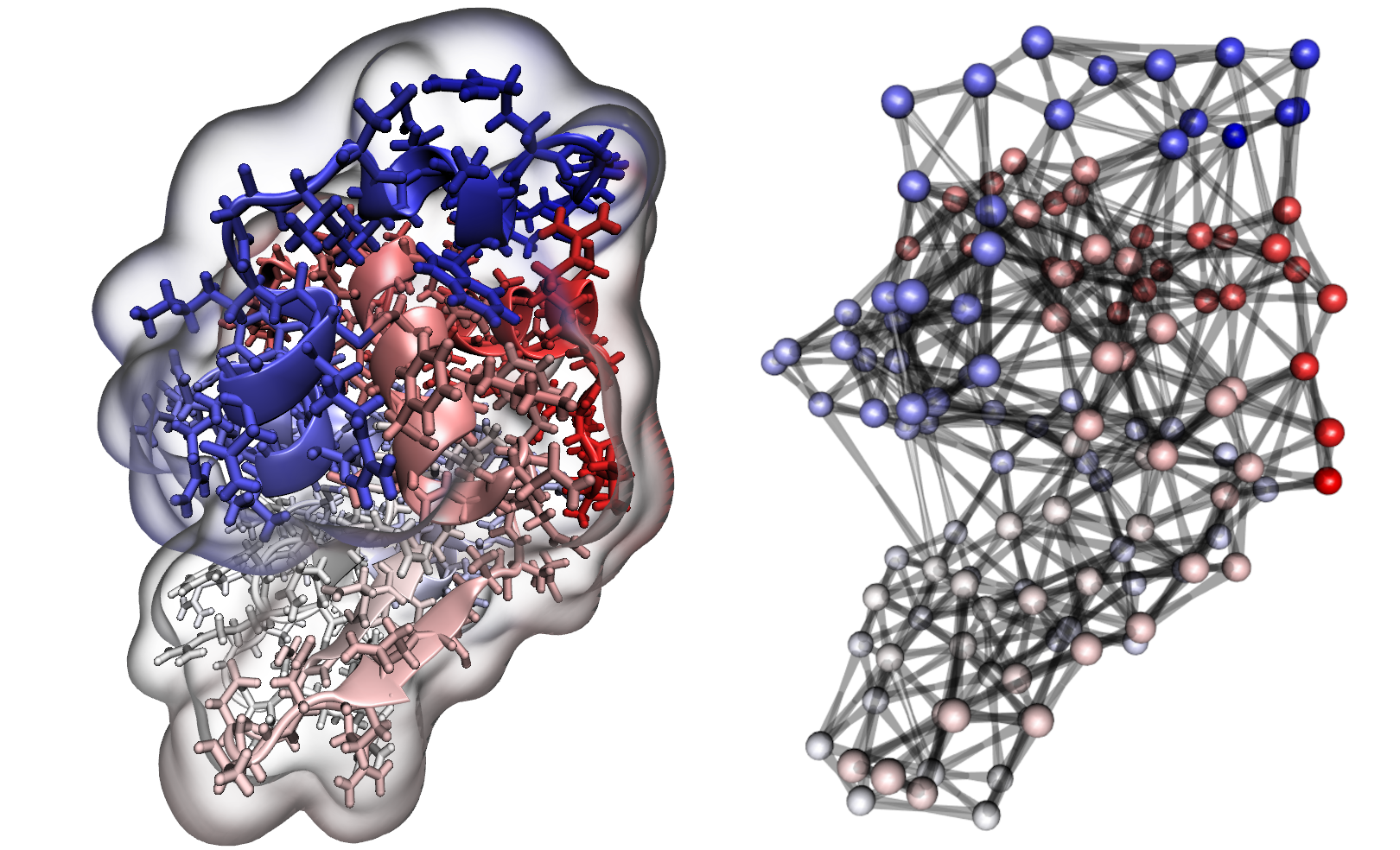}
\caption{3D representations of lysozyme (PDB 1AKI \citep{Artymiuk:a20987}). (Left) Molecular representation, showing backbone (ribbon), side chains, and surface; residues colored by index.  (Right) PSN representation in similar orientation, with vertices positioned by C$\alpha$ coordinates and colored by index. \label{pic:lysozyme_3d}}
\end{figure}

Here, we apply this notion to observed variation in crystal structures of hen egg-white lysozyme (a widely used reference protein in biophysical research).  Lysozyme (N-acetylmuramide glycanhydrolase), is an enzymatic antimicrobial agent produced as part of the innate immune system.  A glycoside hydrolase, lysozyme attacks polysaccharides within bacterial cell walls, compromising their integrity and ultimately causing cell lysis; as such, it is produced in large quantities in settings where bacterial growth must be discouraged (e.g., eggs, tears, milk).  Our data consist of network representations of 66 independently solved lysozyme structures, each of which is formed from 129 residues (i.e. amino acids) constituting the main chain of wild type hen egg-white lysozyme (residues 19-147 of Uniprot B8YK79).  Atomistic protein structures were obtained from the Protein Data Bank (PDB; \emph{https://www.rcsb.org/pdb/home/home.do}), with the search query limited to X-ray crystallography structures containing only the 129-residue main chain with no modified or substituted residues, missing residues, ligands, or other complexes.  Where more than one distinct conformation appeared in the asymmetric unit, each was isolated and treated as a separate conformation for purposes of analysis.  Each isolated protein structure was protonated using REDUCE \citep{word.et.al:jmb:1999}, with the resulting coordinates employed to generate a residue-level protein structure network (i.e. an undirected adjacency matrix of size $129 \times 129$) according to the pairwise distances among residues -- any pair of residues is considered to be adjacent if they contain respective atoms that are closer together than 1.2 times the sum of their respective van der Waals radii.  Two representative lysozyme structure networks are displayed in Fig.~\ref{pic:lysozyme}; while the conformations are very similar, they do show subtle differences (compare e.g., the residues in the top right). A 3D molecular structure of lysozyme is shown in Fig.~\ref{pic:lysozyme_3d}, together with the equivalent protein structure network (PSN).

\subsubsection{Model Specification}

\paragraph{Model terms:} Our model specification includes three categories of effects: covariates relating to residue properties that enhance or inhibit interaction; ``contextual'' covariates relating to the overall fold of the protein; and dependencies among contacts arising from steric and other effects.  Beginning with the first group, we add a Coulomb-like term for interactions based on nominal residue charge, \texttt{ChargeMatch-edgecov}, coded as 1 for pairs with complementary charges, -1 for pairs with non-complementary charges (i.e., positive/positive or negative/negative), and 0 otherwise.  We include two terms for Polar/Polar (\texttt{PolPol-edgecov}) and Nonpolar/Nonpolar (\texttt{NPolNPol-edgecov}) residue pairs, respectively, accounting for the fact that the two affinities are non-identical.  To account for the distinctive interaction patterns of aromatic residues, we include an overall effect for interaction by aromatics (\texttt{Aromatic-nodecov}) as well as an effect for pairwise interactions among aromatic residues \emph{per se} (referred to mnemonically as \texttt{PiStack-edgecov}).  Finally, we account for the greater contact potential of larger residues by incorporating a term for residue surface area (\texttt{SurfaceArea-nodecov}). 

 With respect to the second class of effects, we first observe that distance along the protein backbone is an important predictor of interaction, and we hence include the logged backbone distance as an edgewise covariate (\texttt{logBBDist-edgecov}); separately, we also incorporate adjacency along the backbone as a support constraint (reflecting the fact that each residue is covalently bound to its backbone neighbors).  Because our objective is to model variation in folded lysozyme structures (and not predicting the fold \emph{de novo}), we incorporate an effect for the average distances among residues.  Specifically, we encode the log of the mean distance between alpha carbons (C$\alpha$) for every residue pair (taken over all structures) as an edge covariate (\texttt{logMeanDist-edgecov}), expressing the intuition that residues that are on average spatially proximate in folded lysozyme are more likely to be adjacent in any particular structure.  To account for the fact that surface residues have solvent and/or crystal contacts that are not captured by the structure (resulting in a lower mean degree within the PSN), we also include the mean C$\alpha$ distance from the coordinate center as a nodal covariate (\texttt{meanCADist-nodecov}).  To adjust for differences in the ability of larger or bulkier residues to form contacts at longer C$\alpha$ distances, we also add respective product terms (i.e., interaction effects in a statistical rather than relational sense) between the aromatic and surface area statistics and the log C$\alpha$ distances (\texttt{logMeanDistAro-edgecov} and \texttt{logMeanDistSurf-edgecov}).

Finally, we consider terms relating to the interdependence among contacts.  To model the fact that each of a residue's existing contacts increases the difficulty of forming new contacts, we include a 2-star term (\texttt{2-stars}); likewise, we include a triangle term (\texttt{Triangles}) to account for the increasing difficulty of forming large cliques.  (While both such terms are rarely used in e.g. social network settings due to their propensity to produce degenerate models when their associated coefficients are positive, these terms can be important for capturing geometric constraints in physical systems; since the associated coefficients are generally negative in these cases, they do not lead to runaway clique formation.)  Although large cliques are strongly suppressed by packing constraints, PSNs \emph{are} however highly triangulated.  We thus combine the (hypothesized negative) triangle term with a GWESP term (here using a decay parameter of 0.8 identified by a pilot fit to a single graph).
 
\paragraph{Prior specification:} To specify the prior for conjugate MAP, we begin with the approximation that the mean degree for a fully buried core residue will be approximately 12 (based on a standard sphere packing approximation; see e.g. \citep{hales2005proof}).  In practice, however, many potential contacts are ``lost'' due to residues' not being completely surrounded by other residues (i.e., on the surface).  To approximate the fraction of possible contacts that are ``lost'' in this way, we begin by approximating the expected surface area of the protein that would be used for residue-residue contacts if all residues were buried; paradoxically, this is the surface area of the fully unfolded protein.  \citet{miller.et.al:jmb:1987} show that the empirical model
\[
A_u \approx 1.48 M + 21
\]
provides an excellent approximation to the unfolded surface area of monomeric proteins (where $A_u$ is the surface area in squared Angstroms, and $M$ is the molecular mass in Daltons).  For the surface area of a folded protein, they likewise report the model
\[
A_f \approx 6.3 M^{0.73}
\]
(with the same units as above).  We may approximate the fraction of possible contacts ``lost'' to solvent in the folded protein as $A_f/A_u$, and thus approximate the expected degree by
\[
\bar{d} \approx 12(1-\tfrac{A_f}{A_u}).
\]
For lysozyme, we have $M=14.3$kDa, giving us $A_f \approx 6803.554$ {\AA}$^2$, $A_u \approx 21185$ {\AA}$^2$, and $\bar{d} \approx 8.15$ (i.e., about 32\% of potential residue contacts are predicted to be lost).  Although obtained entirely via \emph{a priori} considerations, we note that this expected degree is quite close to the observed degree for the lysozyme structures in our sample (8.32), suggesting that it is indeed a reasonable choice.  To obtain $\bar{\tau}$, we simulate 1000 conditional Bernoulli graph draws with mean degree $\bar{d}$, subject to the constraint that all backbone-adjacent residues are tied, and take $\bar{\tau}$ equal to the means of the sufficient statistics for the sample.  

To set the prior weight ($n_0$, and hence $\delta$), we observe that our prior information is fairly vague, and we would want the data to outweigh the prior even for a single graph observation.  We thus set $n_0=0.1$, making the prior weight equivalent to one tenth of a single graph observation.  For our data set, with $m=66$, this implies a net prior weight of $\delta \approx 0.0015$.

\subsubsection{Results}

We perform conjugate MAP inference for the pooled ERGM model on the 66 lysozyme PSNs, using the above-specified model; estimation was performed using \textbf{ergm} under default settings incorporating the backbone-adjacency support constraint.  The resulting parameter estimates are provided in Table \ref{tb:lysozyme_analysis_table}. The model parameters can be interpreted based on the conditional log-odds of an edge between two nodes $i$ and $j$, bearing in mind that many effects are necessarily simultaneous.  For example, while the coefficient for the edges term is positive, it should be interpreted in the context of both mean spatial distances and sequence distances between residues. For example, the log mean distance between the C$\alpha$s of residue $4$ and residue $9$ is 2.143, and their log backbone distance is $\log(5)=1.6$.  Ignoring all other effects, then the conditional probability of $Y_{4,9} = 1$ based solely on these three terms would be $[1+\exp[-(34.213 -19.173(2.143)+0.293(1.6))]]^{-1} \approx 0.002$, indicating a low conditional probability of observing an edge; in practice, of course, all terms contribute simultaneously.  

\begin{table}
\begin{center}
\caption{Conjugate Bayesian analysis of Lysozyme structure networks \label{tb:lysozyme_analysis_table}}
\begin{tabular}{lrrr}
\hline\hline
Term & Estimate & (s.d.) &  $95\%$ credible interval \\
\hline
\texttt{Edges}                   & 34.213 & (0.0309) & (34.152, 34.274) \\ 
\texttt{ChargeMatch-edgecov}     & 0.229 & (0.0654) & (0.101, 0.357) \\ 
\texttt{NPolNPol-edgecov}        & 0.760 & (0.0296) & (0.702, 0.818) \\ 
\texttt{PolPol-edgecov}          & 0.186 & (0.0319) & (0.124, 0.249) \\ 
\texttt{Aromatic-nodecov}        & -1.091 & (0.0122) & (-1.115, -1.067) \\ 
\texttt{PiStack-edgecov}         & -0.752 & (0.0849) & (-0.918, -0.585) \\ 
\texttt{SurfaceArea-nodecov}     & -0.040 & (0.0006) & (-0.041, -0.038) \\ 
\texttt{logBBDist-edgecov}       & 0.293 & (0.0110) & (0.272, 0.315) \\ 
\texttt{meanCADist-nodecov}      & -0.040 & (0.0028) & (-0.045, -0.035) \\ 
\texttt{logMeanDist-edgecov}     & -19.173 & (0.0643) & (-19.299, -19.047) \\ 
\texttt{logMeanDistSurf-edgecov} & 0.024 & (0.0003) & (0.023, 0.024) \\ 
\texttt{logMeanDistAro-edgecov}  & 0.646 & (0.0128) & (0.621, 0.671) \\ 
\texttt{GWESP}, $\phi=0.8$       & 1.444 & (0.0385) & (1.368, 1.519) \\ 
\texttt{2-stars}                 & -0.046 & (0.0058) & (-0.057, -0.035) \\ 
\texttt{Triangles}               & -0.829 & (0.0253) & (-0.879, -0.780) \\ 
\hline
\hline
\end{tabular}
\end{center}
\end{table}

As Table~\ref{tb:lysozyme_analysis_table} shows, all three types of mechanisms play a role in predicting lysozyme network structure.  Electrostatic and polar effects act as expected, with complementary charges increasing conditional tie probability and homophily for non-polar residues; although the posterior strongly favors homophily among polar residues, this effect is notably weaker than for the non-polar case.  

Aromatic residues at first blush seem to have a lower baseline contact probability (with an additional negative effect for $\pi$-stacking), but these ``intercept effects'' must be weighed against the reduction in the C$\alpha$ distance penalty for these residues.  Let $i$ and $j$ be residues $d$ {\AA} apart, such that $i$ is aromatic and $j$ is not.  Then the total effect of the Aromatic, $\pi$-stacking, and Aromatic distance effect terms on the conditional log odds of an $i,j$ edge is approximately $-1.091+0.646 \log d$; this enhances tie probability for $d> 5.4${\AA}, only suppressing it at close range.  Likewise, for aromatic-aromatic pairs, the corresponding total effect is approximately $-1.091-0.752+2(0.646) \log d = -1.843+1.292 \log d$, which becomes favorable for $d> 4.2${\AA}.  We would expect mixing among aromatic residues to be favored on physical grounds, and indeed this is true for residue pairs beyond 3.2{\AA}.  Overall, we thus see that interactions with aromatic residues are generally favorable (with aromatic-aromatic mixing especially favorable) except at very close range, with these residues particularly likely to interact with other residues over longer distances.  Very short-range interactions are somewhat hindered for these residues, however, plausibly due to steric effects.

A similar effect is seen for residue size, with surface area having a negative main effect combined with a greater propensity for longer-range interaction; for residues $i$ and $j$ with respective surface areas $s_i$ {\AA$^2$} and $s_j$ {\AA$^2$} at distance $d$ {\AA}, the total effect of the surface area terms on conditional log odds is $-0.04 (s_i+s_j) + 0.024 (s_i + s_j) \log d$, which becomes positive for $d>5.3$ {\AA}.  For reference, the mean non-covalent nearest neighbor distance is approximately 3.8 {\AA}, and the second-nearest is approximately 5.1 {\AA}, so bulk is a positive interaction predictor for the vast majority of potential interactions.  The minor inhibitory effect at very small distances, like that of aromaticity, may reflect steric hindrance.  

Similar subtlety is seen in the case of the mildly positive effect of backbone distance - net of spatial distance - likely reflecting the tendency of the backbone to fold back on itself (creating strong bridges between parts of the protein that are distinct in sequence space).  Note that, marginally, we find that contact probabilities fall off as roughly $\mathrm{BBDist}^{-5/4}$, so this softening effect should not be confused with a net tendency for tie probability to increase with backbone distance.  Rather, we find that when sequence-distant residues happen to be spatially proximate, they are particularly likely to be in contact.  Less nuance is needed to interpret the effect of distance from the origin, or of the mean C$\alpha$ distance between residues: both inhibit contact.  The latter effect is, as has been observed, very large, in keeping with the constraints of a folded protein.  Finally, we observe that net of everything else, existing contacts have an inhibitory effect of new ones (the negative 2-star parameter, cliques are strongly suppressed (negative triangle parameter), and there is an overall tendency towards triangulation net of clique suppression (positive GWESP parameter).

\begin{figure}
\centering
\includegraphics[width=5.5in]{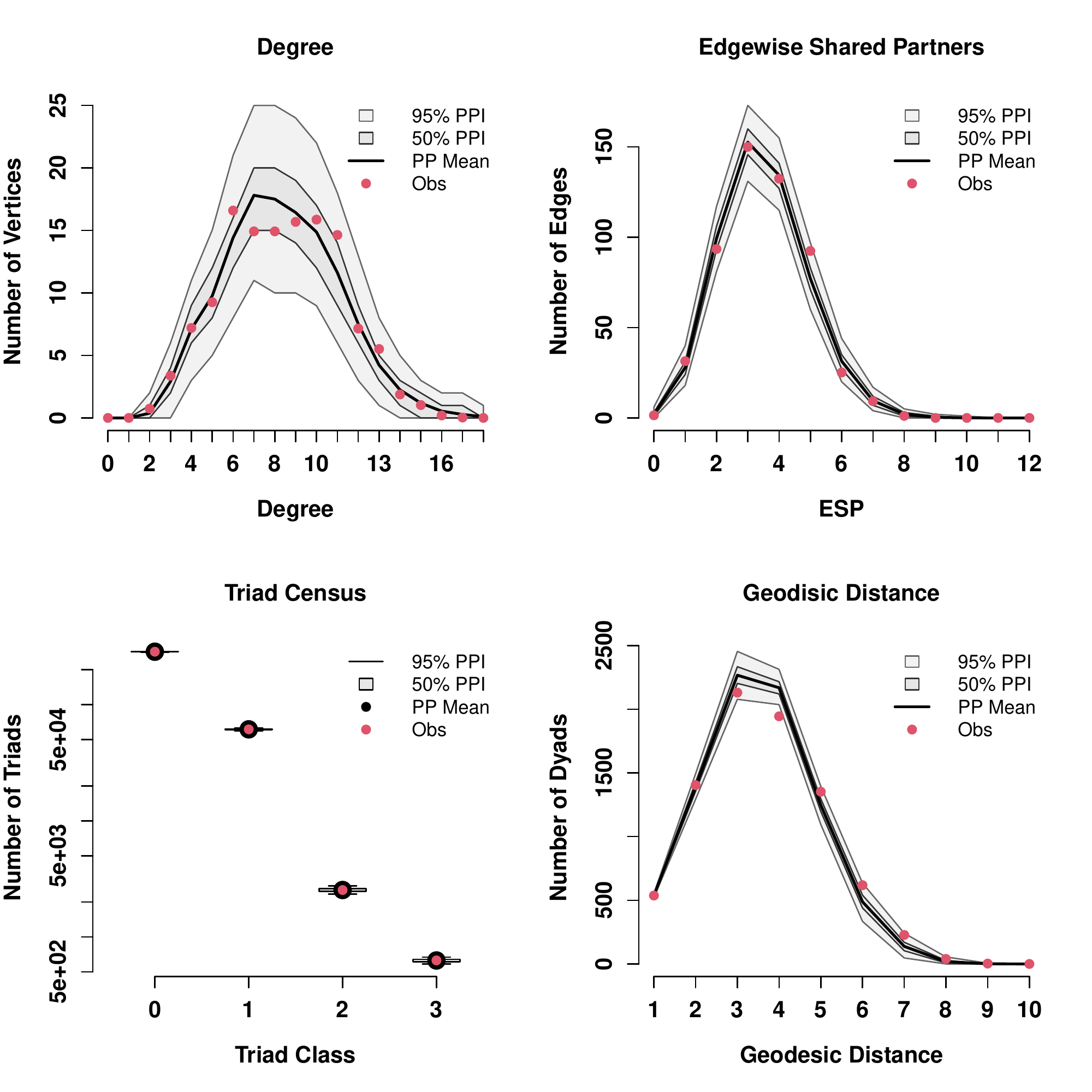}
\caption{Model adequacy checks for the pooled lysozyme model; shaded areas/boxes show posterior predictive intervals, while red points indicate observed mean values.  The lysozyme model successfully recapitulates a range of structural features. \label{pic:lysozyme_gof}}
\end{figure}

As a model adequacy check, we take 1000 draws from the posterior predictive distribution (based on the Laplace approximation), comparing the distribution of several standard structural properties (degree distribution, ESP distribution, geodesic distance distribution, and triad census) with the observed data means.  The result is shown in Fig.~\ref{pic:lysozyme_gof}.  As can be seen, the model is able to recapitulate all of the above features, indicating that it does a reasonable job of capturing the basic structural properties of the lysozyme networks.

\subsubsection{Reproducing Structural Variability}

As noted above, one potential use for ERGM analysis of protein structures is to characterize variability, and to identify dimensions of structural variation that may be imperfectly constrained by available data.  Here, we simulate draws from the fitted lysozyme model and examine their range of variation with respect to four basic graph-level indices (GLIs) found by \citet{unhelkar.et.al:bba:2017} to distinguish protein structures. These are:
\begin{itemize}
\item \emph{Transitivity} \citep{wasserman1994social} -- a standard measure of triadic closure in network analysis, transitivity reflects the compactness of a PSN in the sense that higher levels of transitivity are associated with the structures that are closely and uniformly packed. 
\item Standard deviation of \emph{degree distribution} -- a measure of the level of heterogeneity in local packing around chemical groups.
\item Standard deviation of the \emph{core number} \citep{seidman1983network} -- an indicator of the degree of heterogeneity in structural cohesion, which distinguishes between highly organized structures and structures that combine rigidly and loosely bound regions.
\item Standard deviation of \emph{M-eccentricity} -- the idea of \emph{M-eccentricity} stems from \emph{eccentricity}\citep{west2001introduction}, and was introduced in the context of PSN analysis by \citep{unhelkar.et.al:bba:2017}. The M-eccentricity of a vertex is the mean distance from that vertex to all other vertices; vertices with low M-eccentricity are more centrally located, while those with high M-eccentricity are peripheral to the graph structure. Thus the standard deviation of M-eccentricity distinguishes between uniformly globular structures and structures with deformations or other elongations. 
\end{itemize}

\begin{figure}
\centering
\includegraphics[width=5in]{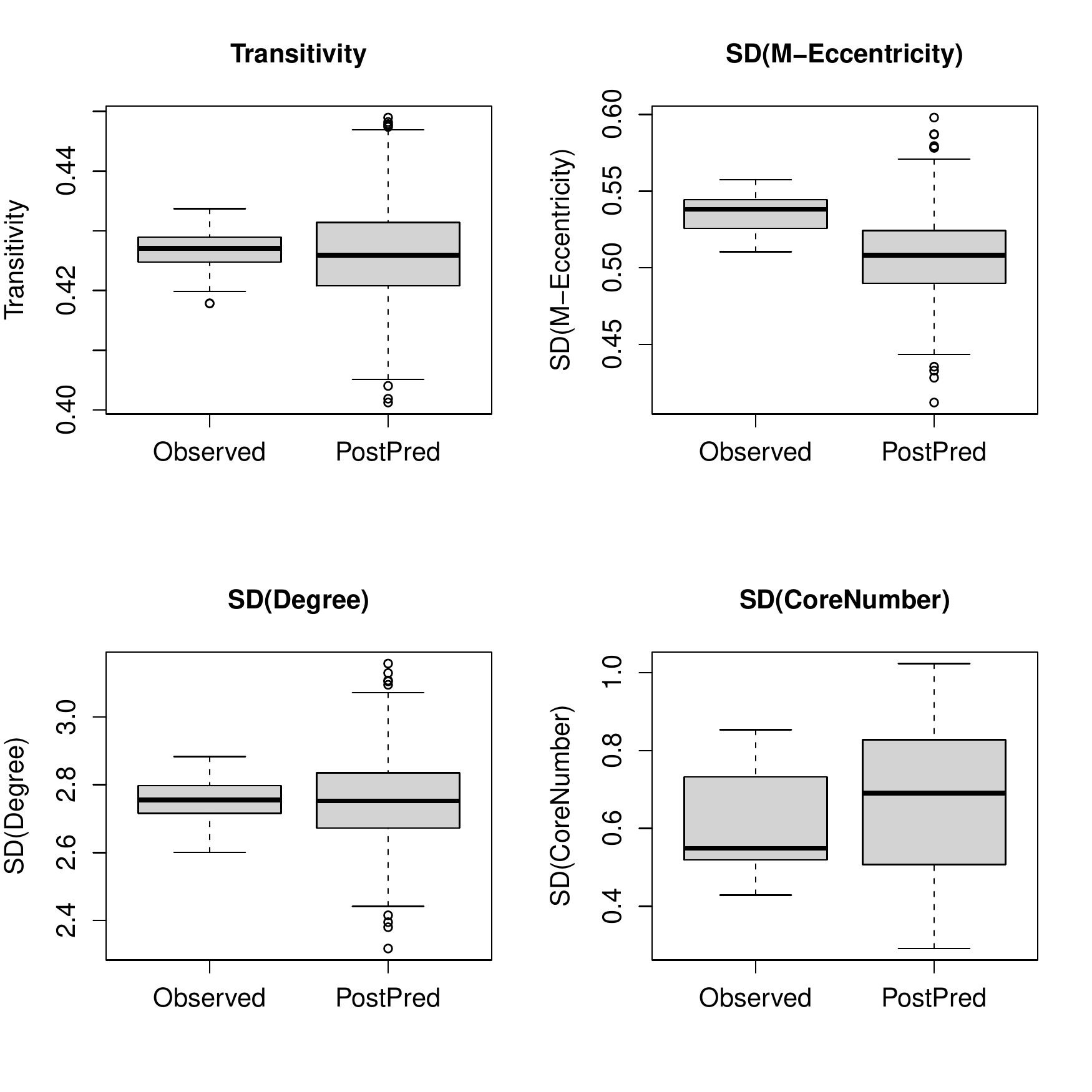}
\caption{Variation in graph-level indices for lysozyme PSNs, observed versus posterior predictive distributions; simulated networks cover the observed GLI range, but also account for predictive uncertainty. \label{pic:lysozyme_gli}}
\end{figure}

Fig.\ref{pic:lysozyme_gof} shows the distribution of the above GLI values for the observed lysozyme networks and for posterior predictive draws from the pooled ERGM; GLIs were calculated using the \textbf{sna} library \citep{butts2008social}.  For each of the GLI distributions, we can see that the posterior predictives cover the observed distributions, while being somewhat wider (reflecting posterior uncertainty).  Such distributions have potential uses for e.g. statistical comparison of protein families or variants from pooled crystallographic data, where accounting for uncertainty in the distribution of structural properties is an important consideration.  

\section{Conclusion}

We have presented a highly scalable approach for modeling multiple network observations with ERGMs, under both frequentist and Bayesian paradigms, utilizing basic exponential family properties to perform pooling and/or Bayesian updating entirely within the mean value space.  This allows us to perform inference on arbitrary numbers of graphs at no additional cost, and to perform Bayesian inference at the same cost as maximum likelihood estimation.  Moreover, by mapping the inferential problem to a problem involving a single network, it is possible to perform both pooled and Bayesian inference with standard software packages designed for single-network applications, without resorting to techniques like graph aggregation with structural zeros that add complexity and computational cost.  Simulation experiments suggest that the frequentist properties of the pooling procedure are quite good (with minimal bias and good calibration with even small sample sizes), and conjugate-prior MAP inference yields well-behaved interpolation between prior parameters and the MLE.  Conjugate-prior MAP estimates with a simple default prior were also found to have good frequentist properties for a range of diffuse prior weights, suggesting its value as a simple tool for regularized inference (with the most important use case being settings where the MLE does not exist due to the convex hull problem).  Although this work focused on a specific choice of default prior that is analogous to a zero vector in the natural parameter space (with the exception of the edge parameter which is corrected for prior density) - a natural analog to the zero-centered priors used in existing strategies for Bayesian ERGM inference - the fact that the conjugate prior is specified in the mean value space (i.e., the space of graph statistics) makes it particularly easy to specify informative alternatives based on e.g. prior data sets.  

We demonstrated the applicability of our inferential scheme with two applications, specifically to brain functional connectivity networks and to protein structure networks.  In both cases, the ability to quickly and easily pool network data without additional computational cost, and to easily use either Bayesian or frequentist inference, facilitates analysis.  We also show how the regularization offered by the use of prior structure makes it possible to include theoretically interesting mixing terms that (because their statistics lie on the convex hull) are problematic under MLE, and how prior substantive information (here, simple empirical models of the properties of monomeric proteins) can be used to create reasonable prior specifications even without existing network data.  

The results shown here were produced using the MCMC MLE estimation strategy used by the \textbf{ergm} package, but the idea can be easily adapted to any other ERGM estimation scheme based on fitting to the sufficient statistics (e.g., contrastive divergence, stochastic approximation, the log partition function scheme of \citet{butts:sm:2007b}, or other forms of gradient descent).  It is not compatible with approximate likelihood methods such as maximum pseudo-likelihood estimation (MPLE) that operate directly on edge variables, although we observe that it is still possible to initialize estimation by MPLE on a single graph from the set and then proceed with methods based on statistics (as was in fact done here), or otherwise use methods such as contrastive divergence that are similar in speed and accuracy.  We do note that one side effect of the high level of statistical precision obtainable from pooled network models is that \emph{de facto} accuracy eventually begins to depend more on numerical error than statistical uncertainty.  While we find that calibration remains good for the range of data sizes considered in our simulation study, precise inference for very large collections of networks may require greater attention to numerical stability than is necessary for conventional ERGM inference.  Efficient high-precision algorithms for pooled models in the large-$m$ regime would seem to be an important problem for future work

\section*{Appendix A}

For convenience, we here provide the definitions of a number of common families of ERGM terms used in this paper.

\texttt{Edges} \newline

\begin{equation*}
\label{eq:ergm_edges}
t_e(y) = \sum_{i<i} y_{ij}
\end{equation*}

This term counts the number of edges within the graph; since the graphs used here are all undirected, we show the undirected version.  The edge term acts as a \emph{de facto} intercept for the model, setting the baseline tie probability, and is central in fixing the expected density of the network.\newline

\texttt{Geometrically weighted edgewise shared partners (GWESP)} \newline
\begin{equation*}
\label{eq:ergm_gwesp}
t_g(y, \phi) = e^{\phi} \sum_{k=1}^{n-2} \left\{ 1 - (1-e^{-\phi})^{k}   \right\} EP_{k}(y)
\end{equation*}

where $EP_{k}(y)$ is the number of connected pairs that have exactly $k$ common neighbors, which is a measure of local clustering in network research. The decay parameter $\phi$ controls the relative contribution of $EP_{k}(y)$ to the GWESP statistic. A positive, large coefficient on GWESP term tends to concentrate more probability mass on the graphs with more local clustering.\newline


\texttt{Nodematch} \newline

Given a categorical covariate $x$ defined for each node in the graph, a nodematch term counts the total number of edges between nodes sharing same value on $x$,



\begin{equation*}
\label{eq:ergm_nodematch_sex}
t_{H}(y;x) = \sum_{i < j} y_{ij}\mathbbm{1}_{ \{x_{i} = x_{j}\} } 
\end{equation*}

\texttt{Nodemix} \newline

Given a categorical covariate $x$ defined for each node in the graph, a nodemix term counts the total number of edges between nodes of level $k$ and level $l$ on $x$, 

\begin{equation*}
\label{eq:ergm_nodemix}
t_M(y;x) = \sum_{i < j} y_{ij}\mathbbm{1}_{ \{ (x_{i},x_{j}) = (k,l) \  \text{or} \ (x_{i},x_{j}) = (l,k)\} } . 
\end{equation*}

The parameter associated with a $k,l$ nodemix statistic inhibits or enhances the rate at which nodes with respective values $k$ and $l$ on $x$ interact with one another; negative values hinder interaction, while positive values enhance it.\newline

\texttt{Nodecov} \newline

Given a covariate $x$ defined for each node in the graph, a nodecov term is defined as follows:

\begin{equation*}
\label{eq:ergm_nodecov}
t_{NC}(y;x) = \sum_{i<j} y_{ij}(x_{i} + x_j)
\end{equation*}

This statistic captures the degree of association between nodes' values of $x$ and their net tendency to send or receive ties. \newline

\texttt{Edgecov} \newline

Given a covariate $x$ defined for each pair of nodes in the graph, an edgecov term is defined as follows:

\begin{equation*}
\label{eq:ergm_edgecov}
t_{EC}(y;x) = \sum_{i<j} y_{ij}x_{ij}
\end{equation*}

This statistic can be viewed as the product-moment of the elements of the adjacency matrix with the elements of $x$, and as such captures the degree of association between dyads' values on $x$ and the presence or absence of edges. \newline

\texttt{2-stars} \newline

 \begin{equation*}
 \label{eq:ergm_2star}
 t_{2S}(y) = \sum_{i<j<k} y_{ik}y_{jk}
 \end{equation*}

 This statistic counts the number of 2-star configurations within the graph (i.e., a subgraph containing node with two simultaneous partners).  The associated parameter can be viewed as indicating the effect of each of a node's current edges on the conditional log odds of a specific other edge involving that node being present; when negative, in particular, it can be viewed as a first-order model for hindrance (in which each existing edge makes it more difficult to acquire new edges). \newline

\texttt{Triangles} \newline

 \begin{equation*}
 \label{eq:ergm_triangle}
 t_T(y) = \sum_{i<j<k} y_{ik}y_{jk}y_{ij}
 \end{equation*}

 This statistic counts the number of triangles with in the graph (i.e., complete subgraphs of order three).  The associated parameter can be viewed as indicating the effect of each of a node pair's shared partners on the conditional log odds of being tied.  Although prone to producing runaway clique formation when associated with a positive parameter (the famous ``density explosion'' path to degeneracy), negative triangle parameters can serve to inhibit the formation of large cliques (since every clique of order $k$ contains on the order of $k^3$ triangles) and can serve as a model for packing constraints.  \newline

\texttt{graphletCount} \newline

This statistic counts the number of times that a specific graphlet appears in a network, and can be added to ERGM via \texttt{graphletCount} command (see \citep{nebil2015ergm} for more details about its definition and implementation) \newline

\section*{Acknowledgments}
This research was supported in part by NIH award 1R01GM144964-01, NASA award 80NSSC20K0620, NSF awards SES-1826589 and DMS-1361425, and ARO award \#W911NF-14-1-0552 to C.T.B. 

\bibliography{final.bib}

\begin{thebibliography}{100}

\bibitem{smith2016ethnic}
Smith S, Van~Tubergen F, Maas I, McFarland DA.
\newblock Ethnic composition and friendship segregation: Differential effects
  for adolescent natives and immigrants.
\newblock American Journal of Sociology. 2016;121(4):1223--1272.

\bibitem{cross.et.al:sn:2001}
Cross R, Borgatti SP, Parker A.
\newblock Beyond Answers: Dimensions of the Advice Network.
\newblock Social Networks. 2001;23(3):215--235.

\bibitem{saul2007exploring}
Saul ZM, Filkov V.
\newblock Exploring biological network structure using exponential random graph
  models.
\newblock Bioinformatics. 2007;23(19):2604--2611.

\bibitem{SAINTBEAT2015458}
Saint-Béat B, Baird D, Asmus H, Asmus R, Bacher C, Pacella SR, et~al.
\newblock Trophic networks: How do theories link ecosystem structure and
  functioning to stability properties? A review.
\newblock Ecological Indicators. 2015;52:458--471.
\newblock doi:{https://doi.org/10.1016/j.ecolind.2014.12.017}.

\bibitem{delmas.et.al:br:2019}
Delmas E, Besson M, Brice MH, Burkle LA, Dalla~Riva GV, Fortin MJ, et~al.
\newblock Analysing ecological networks of species interactions.
\newblock Biological Reviews. 2019;94(1):16--36.
\newblock doi:{https://doi.org/10.1111/brv.12433}.

\bibitem{krause.et.al:bk:2015}
Krause J, James R, Franks D, Croft D, editors.
\newblock Animal Social Networks.
\newblock Oxford: Oxford University Press; 2015.

\bibitem{cross.et.al:bio:2020}
Cross TJ, Takahashi GR, Diessner EM, Crosby MG, Farahmand V, Zhuang S, et~al.
\newblock Sequence Characterization and Molecular Modeling of Clinically
  Relevant Variants of the {SARS-CoV-2} Main Protease.
\newblock Biochemistry. 2020;9(39):3741--3756.
\newblock doi:{10.1021/acs.biochem.0c00462}.

\bibitem{grazioli.et.al:jpcB:2019}
Grazioli G, Yu Y, Unhelkar MH, Martin RW, Butts CT.
\newblock Network-based Classification and Modeling of Amyloid Fibrils.
\newblock Journal of Physical Chemistry, B. 2019;123(26):5452--5462.
\newblock doi:{10.1021/acs.jpcb.9b03494}.

\bibitem{cook.et.al:n:2019}
Cook SJ, Jarrell TA, Brittin CA, Wang Y, Bloniarz AE, A YM, et~al.
\newblock Whole-animal connectomes of both \emph{{C}aenorhabditis elegans}
  sexes.
\newblock Nature. 2019;571:63--–71.

\bibitem{kolaczyk:bk:2009}
Kolaczyk ED.
\newblock Statistical Analysis of Network Data: Methods and Models.
\newblock New York: Springer-Verlag; 2009.

\bibitem{snijders2011statistical}
Snijders TA.
\newblock Statistical models for social networks.
\newblock Annual Review of Sociology. 2011;37.

\bibitem{salter2012review}
Salter-Townshend M, White A, Gollini I, Murphy TB.
\newblock Review of statistical network analysis: models, algorithms, and
  software.
\newblock Statistical Analysis and Data Mining: The ASA Data Science Journal.
  2012;5(4):243--264.

\bibitem{lusher.et.al:bk:2012}
Lusher D, Koskinen J, Robins G.
\newblock Exponential Random Graph Models for Social Networks: Theory, Methods,
  and Applications.
\newblock Cambridge: Cambridge University Press; 2012.

\bibitem{wasserman1996logit}
Wasserman S, Pattison P.
\newblock Logit models and logistic regressions for social networks: {I}. An
  introduction to {M}arkov graphs and $p*$.
\newblock Psychometrika. 1996;61(3):401--425.

\bibitem{holland1981exponential}
Holland PW, Leinhardt S.
\newblock An exponential family of probability distributions for directed
  graphs.
\newblock Journal of the american Statistical association. 1981;76(373):33--50.

\bibitem{frank1986markov}
Frank O, Strauss D.
\newblock Markov graphs.
\newblock Journal of the american Statistical association.
  1986;81(395):832--842.

\bibitem{snijders2006new}
Snijders TA, Pattison PE, Robins GL, Handcock MS.
\newblock New specifications for exponential random graph models.
\newblock Sociological Methodology. 2006;36(1):99--153.

\bibitem{pattison.robins:sm:2002}
Pattison PE, Robins GL.
\newblock Neighborhood-Based Models for Social Networks.
\newblock Sociological Methodology. 2002;32:301--337.

\bibitem{hunter2006inference}
Hunter DR, Handcock MS.
\newblock Inference in curved exponential family models for networks.
\newblock Journal of Computational and Graphical Statistics.
  2006;15(3):565--583.

\bibitem{strauss1986general}
Strauss D.
\newblock On a general class of models for interaction.
\newblock SIAM Review. 1986;28(4):513--527.

\bibitem{haggstrom.jonasson:jap:1999}
H\"{a}ggstr\"{o}m O, Jonasson J.
\newblock Phase Transition in the Random Triangle Model.
\newblock Journal of Applied Probability. 1999;36:1101--1115.

\bibitem{handcock:ch:2003}
Handcock MS.
\newblock Statistical Models for Social Networks: Inference and Degeneracy.
\newblock In: Breiger R, Carley KM, Pattison P, editors. Dynamic Social Network
  Modeling and Analysis. Washington, DC: National Academies Press; 2003. p.
  229--240.

\bibitem{rinaldo2009geometry}
Rinaldo A, Fienberg SE, Zhou Y, et~al.
\newblock On the geometry of discrete exponential families with application to
  exponential random graph models.
\newblock Electronic Journal of Statistics. 2009;3:446--484.

\bibitem{schweinberger2011instability}
Schweinberger M.
\newblock Instability, sensitivity, and degeneracy of discrete exponential
  families.
\newblock Journal of the American Statistical Association.
  2011;106(496):1361--1370.

\bibitem{chatterjee.diaconis:aos:2013}
Chatterjee S, Diaconis P.
\newblock Estimating and Understanding Exponential Random Graph Models.
\newblock Annals of Statistics. 2013;41(5):2428--2461.

\bibitem{butts:jms:2019}
Butts CT.
\newblock A Dynamic Process Interpretation of the Sparse {ERGM} Reference
  Model.
\newblock Journal of Mathematical Sociology. 2019;43(1):40--57.

\bibitem{butts:jms:2020b}
Butts CT.
\newblock Phase Transitions in the Edge/Concurrent Vertex Model.
\newblock Journal of Mathematical Sociology. 2020;forthcoming.
\newblock doi:{10.1080/0022250X.2020.1746298}.

\bibitem{koskinen2004bayesian}
Koskinen J.
\newblock Bayesian analysis of exponential random graphs-estimation of
  parameters and model selection.
\newblock Research Report 2004: 2, Department of Statistics, Stockholm
  University; 2004.

\bibitem{caimo2011bayesian}
Caimo A, Friel N.
\newblock Bayesian inference for exponential random graph models.
\newblock Social Networks. 2011;33(1):41--55.

\bibitem{hunter2012computational}
Hunter DR, Krivitsky PN, Schweinberger M.
\newblock Computational statistical methods for social network models.
\newblock Journal of Computational and Graphical Statistics.
  2012;21(4):856--882.

\bibitem{hummel2012improving}
Hummel RM, Hunter DR, Handcock MS.
\newblock Improving simulation-based algorithms for fitting {ERGMs}.
\newblock Journal of Computational and Graphical Statistics.
  2012;21(4):920--939.

\bibitem{krivitsky2012exponential}
Krivitsky PN.
\newblock Exponential-family random graph models for valued networks.
\newblock Electronic Journal of Statistics. 2012;6:1100.

\bibitem{koskinen.et.al:sn:2013}
Koskinen JH, Robins GL, Wang P, Pattison PE.
\newblock {B}ayesian Analysis for Partially Observed Network Data, Missing
  Ties, Attributes and Actors.
\newblock Social Networks. 2013;35(4):514 -- 527.
\newblock doi:{https://doi.org/10.1016/j.socnet.2013.07.003}.

\bibitem{kolaczyk2015question}
Kolaczyk ED, Krivitsky PN.
\newblock On the question of effective sample size in network modeling: an
  asymptotic inquiry.
\newblock Statistical Science. 2015;30(2):184.

\bibitem{schweinberger.et.al:ss:2020}
Schweinberger M, Krivitsky PN, Butts CT, Stewart J.
\newblock Exponential-Family Models of Random Graphs: Inference in Finite-,
  Super-, and Infinite-Population Scenarios.
\newblock Statistical Science. 2020;35(4):627--662.
\newblock doi:{10.1214/19-STS743}.

\bibitem{goodreau2009birds}
Goodreau SM, Kitts JA, Morris M.
\newblock Birds of a feather, or friend of a friend? Using exponential random
  graph models to investigate adolescent social networks.
\newblock Demography. 2009;46(1):103--125.

\bibitem{srivastava2011culture}
Srivastava SB, Banaji MR.
\newblock Culture, cognition, and collaborative networks in organizations.
\newblock American Sociological Review. 2011;76(2):207--233.

\bibitem{cranmer2011inferential}
Cranmer SJ, Desmarais BA.
\newblock Inferential network analysis with exponential random graph models.
\newblock Political Analysis. 2011;19(1):66--86.

\bibitem{welch2011statistical}
Welch D, Bansal S, Hunter DR.
\newblock Statistical inference to advance network models in epidemiology.
\newblock Epidemics. 2011;3(1):38--45.

\bibitem{grazioli.et.al:fmb:2019}
Grazioli G, Martin RW, Butts CT.
\newblock Comparative Exploratory Analysis of Intrinsically Disordered Protein
  Dynamics using Machine Learning and Network Analytic Methods.
\newblock Frontiers in Molecular Biosciences, Biological Modeling and
  Simulation. 2019;6(42).
\newblock doi:{10.3389/fmolb.2019.00042}.

\bibitem{simpson2011exponential}
Simpson SL, Hayasaka S, Laurienti PJ.
\newblock Exponential random graph modeling for complex brain networks.
\newblock PloS ONE. 2011;6(5):e20039.

\bibitem{simpson2012exponential}
Simpson SL, Moussa MN, Laurienti PJ.
\newblock An exponential random graph modeling approach to creating group-based
  representative whole-brain connectivity networks.
\newblock Neuroimage. 2012;60(2):1117--1126.

\bibitem{sinke2016bayesian}
Sinke MR, Dijkhuizen RM, Caimo A, Stam CJ, Otte WM.
\newblock Bayesian exponential random graph modeling of whole-brain structural
  networks across lifespan.
\newblock Neuroimage. 2016;135:79--91.

\bibitem{knecht2008friendship}
Knecht AB.
\newblock Friendship selection and friends' influence. Dynamics of networks and
  actor attributes in early adolescence.
\newblock Utrecht University; 2008.

\bibitem{zijlstra2006multilevel}
Zijlstra BJ, Van~Duijn MA, Snijders TA.
\newblock The multilevel p2 model.
\newblock Methodology. 2006;2(1):42--47.

\bibitem{faust.skvoretz:sm:2002}
Faust K, Skvoretz J.
\newblock Comparing networks across space and time, size and species.
\newblock Sociological Methodology. 2002;32:267--296.

\bibitem{sweet2013hierarchical}
Sweet TM, Thomas AC, Junker BW.
\newblock Hierarchical network models for education research: Hierarchical
  latent space models.
\newblock Journal of Educational and Behavioral Statistics.
  2013;38(3):295--318.

\bibitem{sweet2014hierarchical}
Sweet TM, Thomas AC, Junker BW.
\newblock Hierarchical mixed membership stochastic blockmodels for multiple
  networks and experimental interventions.
\newblock In: Handbook on Mixed Membership Models and their Applications. Boca
  Raton, FL: Chapman \& Hall/CRC; 2014. p. 463--488.

\bibitem{butts.et.al:joss:2012}
Butts CT, Acton RM, Marcum CS.
\newblock Interorganizational Collaboration in the {H}urricane {K}atrina
  Response.
\newblock Journal of Social Structure. 2012;13.

\bibitem{snijders:sm:2001}
Snijders TAB.
\newblock The Statistical Evaluation of Social Network Dynamics.
\newblock Sociological Methodology. 2001;31:361--395.

\bibitem{koskinen.snijders:jspi:2007}
Koskinen JH, Snijders TAB.
\newblock Bayesian inference for dynamic social network data.
\newblock Journal of Statistical Planning and Inference. 2007;137(12):3930 --
  3938.
\newblock doi:{http://dx.doi.org/10.1016/j.jspi.2007.04.011}.

\bibitem{hanneke2010discrete}
Hanneke S, Fu W, Xing EP, et~al.
\newblock Discrete temporal models of social networks.
\newblock Electronic Journal of Statistics. 2010;4:585--605.

\bibitem{desmarais2012statistical}
Desmarais BA, Cranmer SJ.
\newblock Statistical mechanics of networks: Estimation and uncertainty.
\newblock Physica A: Statistical Mechanics and its Applications.
  2012;391(4):1865--1876.

\bibitem{almquist2014bayesian}
Almquist ZW, Butts CT.
\newblock Bayesian analysis of dynamic network regression with joint
  edge/vertex dynamics.
\newblock Bayesian inference in the social and natural sciences New York City,
  NY: John Wiley \& Sons. 2014;.

\bibitem{krivitsky2014separable}
Krivitsky PN, Handcock MS.
\newblock A separable model for dynamic networks.
\newblock Journal of the Royal Statistical Society: Series B (Statistical
  Methodology). 2014;76(1):29--46.

\bibitem{butts:sm:2011a}
Butts CT.
\newblock {B}ayesian Meta-analysis of Social Network Data Via Conditional
  Uniform Graph Quantiles.
\newblock Sociological Methodology. 2011;41(1):257--298.
\newblock doi:{10.1111/j.1467-9531.2011.01240.x}.

\bibitem{koehly.pattison:ch:2005}
Koehly LM, Pattison P.
\newblock Random Graph Models for Social Networks: Multiple Relations or
  Multiple Raters.
\newblock In: Carrington PJ, Scott J, Wasserman S, editors. Models and Methods
  in Social Network Analysis. Cambridge: Cambridge University Press; 2005. p.
  162--191.

\bibitem{STEWART201998}
Stewart J, Schweinberger M, Bojanowski M, Morris M.
\newblock Multilevel network data facilitate statistical inference for curved
  {ERGMs} with geometrically weighted terms.
\newblock Social Networks. 2019;59:98--119.
\newblock doi:{https://doi.org/10.1016/j.socnet.2018.11.003}.

\bibitem{slaughter2016multilevel}
Slaughter AJ, Koehly LM.
\newblock Multilevel models for social networks: hierarchical Bayesian
  approaches to exponential random graph modeling.
\newblock Social Networks. 2016;44:334--345.

\bibitem{VEGAYON2021225}
{Vega Yon} GG, Slaughter A, {de la Haye} K.
\newblock Exponential random graph models for little networks.
\newblock Social Networks. 2021;64:225--238.
\newblock doi:{https://doi.org/10.1016/j.socnet.2020.07.005}.

\bibitem{zemla.austerweil:cbb:2018}
Zemla JC, Austerweil JL.
\newblock Estimating semantic networks of groups and individuals from fluency
  data.
\newblock Computational Brain and Behavior. 2018;1(1):36--58.
\newblock doi:{10.1007/s42113-018-0003-7}.

\bibitem{beskow.carley:pr:2019}
Beskow DM, Carley KM.
\newblock Agent Based Simulation of Bot Disinformation Maneuvers in {T}witter.
\newblock In: Mustafee N, Bae KHG, Lazarova-Molnar S, Rabe M, Szabo C, Haas P,
  et~al., editors. Proceedings of the IEEE 2019 Winter Simulation Conference.
  IEEE; 2019.

\bibitem{geyer1992constrained}
Geyer CJ, Thompson EA.
\newblock Constrained {M}onte {C}arlo maximum likelihood for dependent data.
\newblock Journal of the Royal Statistical Society Series B (Methodological).
  1992; p. 657--699.

\bibitem{hunter2008ergm}
Hunter DR, Handcock MS, Butts CT, Goodreau SM, Morris M.
\newblock ergm: A package to fit, simulate and diagnose exponential-family
  models for networks.
\newblock Journal of statistical software. 2008;24(3):nihpa54860.

\bibitem{snijders2002markov}
Snijders TA.
\newblock {M}arkov chain {M}onte {C}arlo estimation of exponential random graph
  models.
\newblock Journal of Social Structure. 2002;3(2):1--40.

\bibitem{strauss1990pseudolikelihood}
Strauss D, Ikeda M.
\newblock Pseudolikelihood estimation for social networks.
\newblock Journal of the American Statistical Association.
  1990;85(409):204--212.

\bibitem{schmid2017exponential}
Schmid CS, Desmarais BA.
\newblock Exponential Random Graph Models with Big Networks: Maximum
  Pseudolikelihood Estimation and the Parametric Bootstrap.
\newblock arXiv preprint arXiv:170802598. 2017;.

\bibitem{efron:as:1975}
Efron B.
\newblock Defining the Curvature of a Statistical Problem (with Application to
  Second Order Efficiency) (with Discussion).
\newblock Annals of Statistics. 1975;3:1189--1242.

\bibitem{jaynes:bk:1983}
Jaynes ET.
\newblock Papers on Probability, Statistics, and Statistical Physics.
\newblock Reidel: Dordrecht; 1983.

\bibitem{wang:bk:2011}
Wang R. {B}ayesian Inference of Exponential-family Random Graph Modes for
  Social Networks; 2011.
\newblock Unpublished Doctoral Thesis.

\bibitem{diaconis1979conjugate}
Diaconis P, Ylvisaker D, et~al.
\newblock Conjugate priors for exponential families.
\newblock The Annals of statistics. 1979;7(2):269--281.

\bibitem{bernardo2001bayesian}
Bernardo JM, Smith AF. Bayesian theory; 2001.

\bibitem{van2000asymptotic}
Van~der Vaart AW.
\newblock Asymptotic statistics. vol.~3.
\newblock Cambridge University Press; 2000.

\bibitem{tierney1986accurate}
Tierney L, Kadane JB.
\newblock Accurate approximations for posterior moments and marginal densities.
\newblock Journal of the American Statistical Association. 1986;81(393):82--86.

\bibitem{jeffreys1961theory}
Jeffreys H.
\newblock Theory of probability.
\newblock 3rd ed. New York: Oxford University Press; 1961.

\bibitem{hartigan1964invariant}
Hartigan J, et~al.
\newblock Invariant prior distributions.
\newblock The Annals of Mathematical Statistics. 1964;35(2):836--845.

\bibitem{bernardo1979reference}
Bernardo JM.
\newblock Reference posterior distributions for {B}ayesian inference.
\newblock Journal of the Royal Statistical Society Series B (Methodological).
  1979; p. 113--147.

\bibitem{gelman2008weakly}
Gelman A, Jakulin A, Pittau MG, Su YS, et~al.
\newblock A weakly informative default prior distribution for logistic and
  other regression models.
\newblock The Annals of Applied Statistics. 2008;2(4):1360--1383.

\bibitem{rapoport1953spread}
Rapoport A.
\newblock Spread of information through a population with socio-structural
  bias: I. Assumption of transitivity.
\newblock The Bulletin of Mathematical Biophysics. 1953;15(4):523--533.

\bibitem{erdos1959publicationes}
Erdos P, R{\'e}nyi A.
\newblock Publicationes Mathematicae 6.
\newblock In: On random graphs. vol.~1; 1959. p. 290--297.

\bibitem{gilbert1959random}
Gilbert EN.
\newblock Random graphs.
\newblock The Annals of Mathematical Statistics. 1959;30(4):1141--1144.

\bibitem{resnick1997protecting}
Resnick MD, Bearman PS, Blum RW, Bauman KE, Harris KM, Jones J, et~al.
\newblock Protecting adolescents from harm: findings from the {N}ational
  {L}ongitudinal {S}tudy on {A}dolescent {H}ealth.
\newblock JAMA. 1997;278(10):823--832.

\bibitem{team2017r}
Team RC. R: A language and environment for statistical computing. {V}ienna,
  {A}ustria: {R} {F}oundation for {S}tatistical {C}omputing; 2016; 2017.

\bibitem{handcock.et.al:jss:2008}
Handcock MS, Hunter DR, Butts CT, Goodreau SM, Morris M.
\newblock {statnet}: Software Tools for the Representation, Visualization,
  Analysis and Simulation of Network Data.
\newblock Journal of Statistical Software. 2008;24(1):1--11.

\bibitem{butts2008network}
Butts CT, et~al.
\newblock network: a Package for Managing Relational Data in R.
\newblock Journal of Statistical Software. 2008;24(2):1--36.

\bibitem{butts2008social}
Butts CT, et~al.
\newblock Social network analysis with sna.
\newblock Journal of statistical software. 2008;24(6):1--51.

\bibitem{krivitsky2011adjusting}
Krivitsky PN, Handcock MS, Morris M.
\newblock Adjusting for network size and composition effects in
  exponential-family random graph models.
\newblock Statistical Methodology. 2011;8(4):319--339.

\bibitem{simpson2013analyzing}
Simpson SL, Bowman FD, Laurienti PJ.
\newblock Analyzing complex functional brain networks: fusing statistics and
  network science to understand the brain.
\newblock Statistics Surveys. 2013;7:1.

\bibitem{rubinov2010complex}
Rubinov M, Sporns O.
\newblock Complex network measures of brain connectivity: uses and
  interpretations.
\newblock Neuroimage. 2010;52(3):1059--1069.

\bibitem{peiffer2009aging}
Peiffer AM, Hugenschmidt CE, Maldjian JA, Casanova R, Srikanth R, Hayasaka S,
  et~al.
\newblock Aging and the interaction of sensory cortical function and structure.
\newblock Human Brain Mapping. 2009;30(1):228--240.

\bibitem{tzourio2002automated}
Tzourio-Mazoyer N, Landeau B, Papathanassiou D, Crivello F, Etard O, Delcroix
  N, et~al.
\newblock Automated anatomical labeling of activations in {SPM} using a
  macroscopic anatomical parcellation of the {MNI} {MRI} single-subject brain.
\newblock Neuroimage. 2002;15(1):273--289.

\bibitem{hayasaka2010comparison}
Hayasaka S, Laurienti PJ.
\newblock Comparison of characteristics between region-and voxel-based network
  analyses in resting-state {fMRI} data.
\newblock Neuroimage. 2010;50(2):499--508.

\bibitem{venables.ripley:bk:2002}
Venables WN, Ripley BD.
\newblock Modern Applied Statistics with S. Fourth Edition.
\newblock New York: Springer; 2002.
\newblock Available from: \url{http://www.stats.ox.ac.uk/pub/MASS4/}.

\bibitem{daraganova2012networks}
Daraganova G, Pattison P, Koskinen J, Mitchell B, Bill A, Watts M, et~al.
\newblock Networks and geography: Modelling community network structures as the
  outcome of both spatial and network processes.
\newblock Social networks. 2012;34(1):6--17.

\bibitem{nebil2015ergm}
Nebil Y, Fitzhugh SM, Kurant M, Markopoulou A, Butts CT, Prulj N, et~al.
\newblock ergm. graphlets: A Package for ERG Modeling Based on Graphlet
  Statistics.
\newblock Journal of Statistical Software. 2015;65(i12).

\bibitem{amitai2004network}
Amitai G, Shemesh A, Sitbon E, Shklar M, Netanely D, Venger I, et~al.
\newblock Network analysis of protein structures identifies functional
  residues.
\newblock Journal of molecular biology. 2004;344(4):1135--1146.

\bibitem{bode2007network}
B{\"o}de C, Kov{\'a}cs IA, Szalay MS, Palotai R, Korcsm{\'a}ros T, Csermely P.
\newblock Network analysis of protein dynamics.
\newblock Febs Letters. 2007;581(15):2776--2782.

\bibitem{chakrabarty2016naps}
Chakrabarty B, Parekh N.
\newblock {NAPS}: Network analysis of protein structures.
\newblock Nucleic Acids Research. 2016;44(W1):W375--W382.

\bibitem{unhelkar.et.al:bba:2017}
Unhelkar MH, Duong VT, Enendu KN, Kelly JE, Tahir S, Butts CT, et~al.
\newblock Structure Prediction and Network Analysis of Chitinases from the
  {C}ape Sundew, \emph{{D}rosera Capensis}.
\newblock Biochimica et Biophysica Acta - General Subjects.
  2017;1861(3):636--643.

\bibitem{yu.et.al:nsr:2020}
Yu Y, Grazioli G, Unhelkar M, Martin RW, Butts CT.
\newblock Network {H}amiltonian Models Reveal Pathways to Amyloid Fibril
  Formation.
\newblock Nature Scientific Reports. 2020;10:15668.
\newblock doi:{10.1038/s41598-020-72260-8}.

\bibitem{word.et.al:jmb:1999}
Word JM, Lovell SC, Richardson JS, Richardson DC.
\newblock Asparagine and Glutamine: Using Hydrogen Atom Contacts in the Choice
  of Sidechain Amide Orientation.
\newblock Journal of Molecular Biochemistry. 1999;285:1735--1747.

\bibitem{Artymiuk:a20987}
Artymiuk PJ, Blake CCF, Rice DW, Wilson KS.
\newblock {The structures of the monoclinic and orthorhombic forms of hen
  egg-white lysozyme at 6 {\AA} resolution}.
\newblock Acta Crystallographica Section B. 1982;38(3):778--783.
\newblock doi:{10.1107/S0567740882004075}.

\bibitem{Ramin:ha0170}
Ramin M, Shepard W, Fourme R, Kahn R.
\newblock {Multiwavelength anomalous solvent contrast (MASC): derivation of
  envelope structure-factor amplitudes and comparison with model values}.
\newblock Acta Crystallographica Section D. 1999;55(1):157--167.
\newblock doi:{10.1107/S090744499800626X}.

\bibitem{hales2005proof}
Hales TC.
\newblock A proof of the {K}epler conjecture.
\newblock Annals of Mathematics. 2005;162(3):1065--1185.

\bibitem{miller.et.al:jmb:1987}
Miller S, Janin J, Lesk AM, Chothia C.
\newblock Interior and Surface of Monmeric Proteins.
\newblock Journal of Molecular Biology. 1987;196:641--656.

\bibitem{wasserman1994social}
Wasserman S, Faust K.
\newblock Social Network Analysis: Methods and Applications. vol.~8.
\newblock Cambridge University Press; 1994.

\bibitem{seidman1983network}
Seidman SB.
\newblock Network structure and minimum degree.
\newblock Social Networks. 1983;5(3):269--287.

\bibitem{west2001introduction}
West DB, et~al.
\newblock Introduction to Graph Theory. vol.~2.
\newblock Upper Saddle River: Prentice Hall; 2001.

\bibitem{butts:sm:2007b}
Butts CT.
\newblock Models for Generalized Location Systems.
\newblock Sociological Methodology. 2007;37(1):283--348.
\newblock doi:{10.1111/j.1467-9531.2006.00187.x}.

\end{thebibliography}




\end{document}